\documentclass[fleqn,usenatbib]{mnras}
\usepackage[utf8]{inputenc}
\usepackage{graphicx}
\usepackage{amsmath}
\usepackage[dvipsnames]{xcolor}
\usepackage{amssymb}
\usepackage{listings}
\usepackage{xspace}

\usepackage{algorithm}
\usepackage{algpseudocode}

\usepackage{wrapfig}
\usepackage{lscape}
\usepackage{rotating}
\usepackage{epstopdf}

\usepackage[T1]{fontenc}
\usepackage{ae,aecompl}
\newcommand{\healpix}[0]{\textsc{HEALPix}}
\newcommand{\glimpse}[0]{\textsc{Glimpse}}
\newcommand{\nside}[0]{\texttt{NSIDE}}
\newcommand{\redmagic}[0]{redMaGiC}
\DeclareMathOperator{\Imag}{Im}

\def\Sref#1{Sec.~\ref{#1}\xspace}

\usepackage{lineno}

\let\oldequation\equation
\let\oldendequation\endequation

\renewenvironment{equation}
  {\linenomathNonumbers\oldequation}
  {\oldendequation\endlinenomath}

\usepackage{eso-pic}

\AddToShipoutPictureBG*{%
  \AtPageUpperLeft{%
    \hspace{0.73\paperwidth}%
    \raisebox{-4.7\baselineskip}{%
      \makebox[0pt][l]{\textnormal{DES 2019-0489}}
}}}%

\AddToShipoutPictureBG*{%
  \AtPageUpperLeft{%
    \hspace{0.73\paperwidth}%
    \raisebox{-5.7\baselineskip}{%
      \makebox[0pt][l]{\textnormal{FERMILAB-PUB-21-342-AE-SCD}}
}}}%
\title[DES Year 3: weak lensing mass map reconstruction]{Dark Energy Survey Year 3 results: curved-sky weak lensing mass map reconstruction}

\author[N. Jeffrey, M. Gatti et al.]{
\parbox{\textwidth}{
\large{N.~Jeffrey,$^{1,2}$\thanks{E-mail: niall.jeffrey@phys.ens.fr}
M.~Gatti,$^{3,4}$\thanks{E-mail: mgatti@ifae.es}
C.~Chang,$^{5,6}$
L.~Whiteway,$^{2}$
U.~Demirbozan,$^{3}$
A.~Kovacs,$^{7,8}$
G.~Pollina,$^{9}$
D.~Bacon,$^{10}$
N.~Hamaus,$^{9}$
T.~Kacprzak,$^{11}$
% ...
\begin{center} (DES Collaboration) \end{center}
}
\parbox{\textwidth}{ \small
\textit{The authors' affiliations are shown in Appendix~\ref{sec:affiliations}. \\
}}
}}
\date{Accepted 2021 May 11. Received 2021 May 07; in original form 2021 March 26 
}

\pubyear{2021}

\begin{document}

\maketitle

\begin{abstract}
We present reconstructed convergence maps, \textit{mass maps}, from the Dark Energy Survey (DES) third year (Y3) weak gravitational lensing data set. The mass maps are weighted projections of the density field (primarily dark matter) in the foreground of the observed galaxies. We use four reconstruction methods, each is a \textit{maximum a posteriori} estimate with a different model for the prior probability of the map: Kaiser-Squires, null B-mode prior, Gaussian prior, and a sparsity prior. All methods are implemented on the celestial sphere to accommodate the large sky coverage of the DES Y3 data. We compare the methods using realistic $\Lambda$CDM simulations with mock data that are closely matched to the DES Y3 data. We quantify the performance of the methods at the map level and then apply the reconstruction methods to the DES Y3 data, performing tests for systematic error effects. The maps are compared with optical foreground cosmic-web structures and are used to evaluate the lensing signal from cosmic-void profiles. The recovered dark matter map covers the largest sky fraction of any galaxy weak lensing map to date.
\end{abstract}

\begin{keywords}
gravitational lensing: weak -- cosmology: large-scale structure of Universe -- methods: statistical 
\end{keywords}

\section{Introduction}
\label{sec:intro}

Weak gravitational lensing is one of the primary cosmological probes of recent galaxy surveys \citep[for a detailed review of weak lensing see][]{Bartelmann2001,Mandelbaum2018}. By measuring the subtle distortions of galaxy shapes due to the mass distribution between the observed galaxies and us the observers, we are able to place tight constraints on the cosmological model describing the Universe and associated nuisance parameters. In particular, weak lensing most tightly constrains the content of matter in the Universe ($\Omega_{m}$) as well as the level at which matter clusters ($\sigma_{8}$, defined to be the standard deviation of the linear overdensity fluctuations on a $8~ h^{-1}~\mathrm{Mpc}$ scale). Weak lensing also has great potential to constrain dark energy by using galaxy shapes measured at a range of redshifts. In addition to information about the cosmological model describing the Universe, the reconstructed maps of the mass distribution from weak lensing are rich in information about the interaction between galaxies, clusters, and the cosmic web.

The main focus of weak lensing analyses to date has been the measurement of two-point summary statistics such as correlation functions or power spectra \citep{Troxel2017, Hildebrandt2017, Hikage2019, Hamana2020}. A zero-mean Gaussian density field can be statistically completely characterized by its two-point statistics. The methodologies for measuring and modelling these two-point statistics are now relatively well-developed and standard analyses of two-point statistics in weak lensing now take into account several non-trivial systematic effects that were not known a decade ago. These effects include intrinsic alignment (IA), clustering of source galaxies, small-scale modelling of baryonic effects, and uncertainty in photometric redshift calibrations \citep[a detailed review of recent developments in these areas can be found in][]{Mandelbaum2017}. 

In the standard model of cosmology, the initial highly-Gaussian density field becomes increasingly non-Gaussian on small scales through non-linear structure formation. As the techniques for two-point analyses mature, it is natural to ask whether we could extract significantly more information from the same data simply by going to higher order (i.e. non-Gaussian) summary statistics, and whether we understand, at the same level as the two-point statistics, the non-trivial systematic effects in these higher-order statistics. Common higher-order statistics with weak lensing include shear peak statistics \citep{Dietrich2010, Kratochvil2010, Liu2015, Kacprzak2016, Martinet2018, Peel2018, Shan2018, ajani_peaks}, higher moments of the weak lensing convergence \citep{VanWaerbeke2013,Petri2015,Vicinanza2016,Chang2018,Vicinanza2018,Peel2018, Gatti2019}, three-point correlation functions or bispectra \citep{Takada2003, Takada2004, Semboloni2011, Fu2014}, Minkowski functionals \citep{Kratochvil2012,Petri2015,Vicinanza2019,Parroni2020}, and machine-learning methods \citep{Ribli2018, Fluri2018,Fluri2019,jeffrey_lfi}. Many of these have recently been applied to data \citep{Liu2015, Kacprzak2016,Martinet2018,Fluri2019,jeffrey_lfi}, often performing well in terms of cosmological constraints. 

This paper will focus on the key element for many of the methods described above: a weak lensing convergence map, often referred to as a \textit{mass map}. Such a map quantifies the integrated total mass along the line of sight (weighted by a lensing efficiency that peaks roughly half-way between the source and the observer). Two crucial features make a convergence map appealing for extracting higher-order statistics: 1) the map preserves the \textit{phase} information of the mass distribution and 2) the convergence is a scalar field, which can be easier to manipulate/model than a shear field (the latter is closer to what we observe, as explained in \Sref{sec:theory}). Many methods for generating these convergence maps have been proposed; the foundation of most of them is the direct inversion algorithm developed in \citet[][hereafter KS]{Kaiser1993}, a purely analytic solution for converting between shear (the observable) and convergence. Many papers are based on the KS method, including cosmological analyses \citep{VanWaerbeke2013, Vikram2015, Chang2015, Liu2015, Chang2018, Oguri2018}.

The main difficulties associated with the KS method are the treatment of the noise and mask effects.  {In practice, galaxy surveys only observe a part of the sky, and mask out different regions of their sky footprint where the shear field cannot be properly estimated. This usually affects the map-making process, resulting in a poor estimate of the convergence field near masked regions and near the edge of the footprint. Moreover, we can observe only a noisy realization of the shear field, which often leads to a noise-dominated estimate of the convergence field.} Methods more sophisticated than KS were developed to deal with these issues. These include noise modelling and signal priors, either in closed-form \citep{marshall_mass_maps, lanusse_2016, alsing2017cosmological, jeffrey2018, price_maps} -- this is the approach we will take in this work -- or implicitly learned using samples from the prior (e.g. using deep learning \citealt{deeplearning_shirasaki, Jeffrey2020}).  {Many methods have been shown to improve some aspects of the reconstruction of the convergence maps, but ultimately the choice of method depends on the science application of these maps. 

Therefore there is no single comprehensive test for comparative performance between methods; a number of different tests have to be considered.}

One goal of this paper is to present an objective and systematic comparison between several map reconstruction methods using the same set of simulations and data. We present results using the DES Y3 shear catalogue of 100,204,026 galaxies in 4143~deg$^2$. These results highlight expected differences in the maps constructed using the different algorithms and illustrate the advantages or disadvantages of their use in different science cases. We present a comprehensive framework under which most of the convergence map-making methods described previously can be connected and compared. We focus particularly on four methods that span the range of the most popular methods: KS, null B-mode prior, Gaussian prior (Wiener), and halo-model sparsity prior (\glimpse{}). The methods are applied first to a set of DES Y3-like mock galaxy catalogues to demonstrate the performance of each method when the true underlying convergence field is known.

Applying the four methods to the DES Y3 data, we fulfil further goals of performing tests for effects of observational systematic error. We compare the reconstructed weak lensing convergence maps with DES observations of foreground structures; this has further applications for future cosmographic studies and full analyses correlating these maps with cosmological observables (e.g. type Ia supernovae, galaxies and cosmic web structures). Further papers (to follow) will use the maps generated here for cosmology analyses and inference.

The structure of the paper is as follows: in \Sref{sec:theory} we provide the theoretical background for weak gravitational lensing and the framework that connects convergence with observable quantities in a galaxy survey. In \Sref{sec:massmapping} we present a mathematical framework in which the four different mass mapping methods of interest (KS, null B-mode, Wiener, \glimpse{}) are seen to differ only with respect to the priors that are adopted. The data products and simulations used in this work are described in \Sref{sec:data_sims}. In \Sref{sec:sims_maps} we carry out a series of tests on mass maps generated from the four methods and compare them systematically. We then apply the four methods to the DES Y3 data in \Sref{sec:data_maps} and present tests for additional systematic residuals from observational effects. We additionally compare and analyse the maps with observations of foreground structures. We conclude in \Sref{sec:summary}. 

\section{Weak gravitational lensing on the sphere}
\label{sec:theory}

We begin with the gravitational potential $\Phi$ and the matter overdensity field $\delta \equiv \delta\rho / \bar{\rho}$; these real scalar fields on spacetime are related by the Poisson equation

\begin{equation}
\label{eq:poisson}
\nabla^2_r \Phi(t, \boldsymbol{r}) = \frac{3 \Omega_m H_0^2}{2 a(t)} \delta(t, \boldsymbol{r}) \ .
\end{equation}

\noindent Here $t$ is time, $\boldsymbol{r}$ is a comoving spatial coordinate, $\Omega_m$ is the total matter density today, $H_0$ is the Hubble constant today, and $a \equiv 1/(1+z)$ is the scale factor.

Weak gravitational lensing is the small distortion of the shapes of distant galaxies caused by the gravitational warping of spacetime (and hence the distortion of light paths) by mass located between the galaxies and an observer; see \cite{Bartelmann2001} for a comprehensive introduction.

We will parametrize the observer's past lightcone as $(\chi, \theta, \varphi)$ with $\chi$ the comoving radial distance from the observer and $\theta$, $\varphi$ a point on the observer's celestial sphere. The effect of weak lensing can be encapsulated in the \textit{lensing potential}, denoted $\phi$, a real scalar field on the lightcone; its value is related to the gravitational potential $\Phi$ projected along the line of sight:

\begin{equation}
\label{eq:born}
\phi (\chi,\theta,\varphi) = \frac{2}{c^2} \int_0^{\chi} d \chi' \frac{f_K (\chi -\chi')}{f_K (\chi)f_K (\chi')} \Phi (\chi',\theta,\varphi).
\end{equation}
\noindent This equation assumes the Born approximation (the path of integration is not perturbed by the intervening mass). Here the angular distance function $f_K$ is $\sin$, the identity, or $\sinh$ depending on whether the curvature $K$ is positive, zero, or negative.

The radial dependence of $\phi$ in equation~\ref{eq:born} would allow a three-dimensional analysis; however, instead of this, we integrate away the radial dependence using as a weight function the normalised redshift distribution $n(z)$ of source galaxies, obtaining

\begin{equation}
\phi (\theta,\varphi) = \int d \chi \ n(z(\chi)) \ \phi (\chi,\theta,\varphi),
\end{equation}

\noindent a real scalar field on the celestial sphere.

To handle $\phi$ as well as derived quantities we use the formalism of spin-weight functions on the sphere as described in \cite{Castro-PhysRevD.72.023516}. Let $_s Y_{lm}(\theta,\varphi)$ denote the spin-weight $s$ spherical harmonic basis functions. Recall that the covariant derivative $\eth$ increments the spin-weight $s$ while its adjoint $\bar{\eth}$ decrements it; these operators act in a straightforward fashion on the basis functions.

The convergence $\kappa = \kappa_E + i\kappa_B$ (of spin-weight 0 i.e. a scalar) and shear $\gamma = \gamma_1 + i\gamma_2$ (of spin-weight 2) are related to the lensing potential via:

\begin{equation}
\label{eq:kappa}
  \kappa = \frac{1}{4} (\eth \bar{\eth} +\bar{\eth} \eth) \phi ,
\end{equation}

\begin{equation}
\label{eq:gamma}
  \gamma = \frac{1}{2} \eth \eth  \phi .
\end{equation}

\noindent The convergence satisfies

\begin{equation}
\begin{split}
& \kappa(\theta, \phi) = \frac{3 \Omega_m H_0^2}{2 c^2} \times \\
& \ \int_0^{\infty} d \chi \ n(z(\chi)) \ \int_0^{\chi} d \chi' \frac{f_K (\chi') f_K (\chi -\chi')}{f_K (\chi)} \frac{\delta(\chi', \theta, \phi)}{a(\chi')}.
\end{split}
\end{equation}

We now move to harmonic space, obtaining harmonic coefficients $\hat{\phi}_{\ell m}$, $\hat{\kappa}_{\ell m}$ and $\hat{\gamma}_{\ell m}$ for $\phi$, $\kappa$ and $\gamma$ respectively. Here for example:

\begin{equation}
\label{eq:gammaSHexpansion}
\gamma = \sum_{\ell m} \hat{\gamma}_{\ell m} \, _2Y_{\ell m}
\end{equation}

\noindent with

\begin{equation}
\hat{\gamma}_{\ell m}  = \int d\Omega \ \gamma(\theta ,\varphi) \, _2Y_{\ell m}^{*}(\theta ,\varphi) .
\end{equation}

\noindent We can decompose the harmonic coefficients into real and imaginary parts: $\hat{\kappa}_{\ell m} = \hat{\kappa}_{E,\ell m} + i \hat{\kappa}_{B,\ell m}$ and $\hat{\gamma}_{\ell m} = \hat{\gamma}_{E,\ell m} + i \hat{\gamma}_{B,\ell m}$. In harmonic space, equations \ref{eq:kappa} and \ref{eq:gamma} become:

\begin{equation}
\hat{\kappa}_{\ell m}  = - \frac{1}{2} \ell (\ell+1) \hat{\phi}_{\ell m}
\end{equation}

\noindent and

\begin{equation}
\hat{\gamma}_{lm} = \frac{1}{2}\sqrt{(\ell-1)\ell(\ell+1)(\ell+2)}\hat{\phi}_{\ell m} .
\end{equation}

\noindent Thus

\begin{equation}
\label{eq:mass_map_operator}
\hat{\gamma}_{lm} = -\sqrt{\frac{(\ell-1)(\ell+2)}{\ell(\ell+1)}} \hat{\kappa}_{\ell m}.
\end{equation}

\section{Mass map inference}
\label{sec:massmapping}

The formalism introduced in the previous section relates an ideal complex shear field defined on the full celestial sphere $\gamma$ to the convergence field $\kappa$ for a given source redshift distribution. This ideal shear field is full-sky, sampled everywhere, and noise-free. Inferring the unknown convergence field from ellipticity measurements of a finite set of source galaxies in the presence of survey masks and galaxy \textit{shape noise} (discussed below) is the challenge of mass mapping. 

The real and imaginary parts of the shear $\gamma$ are relative to a chosen two dimensional coordinate system. In weak lensing, the observed ellipticity (\cite{Bartelmann2001} equation 4.10) of a galaxy $\epsilon_{\mathrm{obs}}$ is related to the reduced shear $g$ plus the intrinsic ellipticity of the source galaxy $\epsilon_\mathrm{s}$ through

\begin{equation} \label{eq:recuded_shear}
\begin{split}
\epsilon_{\mathrm{obs}} &\approx g + \epsilon_\mathrm{s}, \\
& \ \  \mathrm{where} \ \ g = \frac{\gamma}{1 - \kappa} \ .
\end{split}
\end{equation}

\noindent In the weak lensing limit, the reduced shear is approximately the true shear, $g \approx \gamma $. This allows an observed shear to be defined, $\gamma_{\mathrm{obs}} = \epsilon_{\mathrm{obs}}$; this can be interpreted as a noisy measurement of the true shear that has been degraded by shape noise (caused by the unknown intrinsic ellipticities $\epsilon_\mathrm{s}$ of the observed galaxies):

\begin{equation} \label{eq:shape_noise}
\gamma_{\mathrm{obs}} \approx \gamma +   \epsilon_\mathrm{s} \ .
\end{equation}

\noindent The shape noise is larger than the lensing signal by a factor of  $\mathcal{O}(100)$ per galaxy. It is therefore a dominant source of noise.

In a Bayesian framework we consider the posterior distribution of the convergence $\kappa$ conditional on the observed shear $\gamma$ (here we have dropped the subscript $_{\mathrm{obs}}$ for brevity) and on the model $\mathcal{M}$:

\begin{equation} \label{eq:bayes_theorem}
    p({\kappa} | {\gamma}, \mathcal{M}) = \frac{p({\gamma} | {\kappa}, \mathcal{M}) \ p({\kappa} | \mathcal{M})}{p({\gamma} | \mathcal{M})} \ \ ,
\end{equation}

\noindent where $p({\gamma} | {\kappa}, \mathcal{M})$ is the likelihood (encoding the noise model), $p({\kappa} | \mathcal{M})$ is the prior, and $p({\gamma} | \mathcal{M})$ is the Bayesian evidence.

We formulate all reconstructed convergence $\kappa$ maps as the most probable maps (given our observed data and assumptions); this is the peak of the posterior i.e. the \textit{maximum a posteriori} estimate. From equation~\ref{eq:bayes_theorem} we see that the \textit{maximum a posteriori} estimate is given by

\begin{equation} \label{eq:posterior_optimization}
\hat{\boldsymbol{\kappa}} = \underset{\boldsymbol{\kappa}}{\rm arg \ max} \  \log p(\boldsymbol{\gamma} | \boldsymbol{\kappa}, \mathcal{M}) + \log p(\boldsymbol{\kappa} | \mathcal{M}) \ ,
\end{equation}

\noindent where $\mathcal{M}$ is our model (which in our case changes depending on the chosen prior distribution). Here, the elements of the vectors $\boldsymbol{\kappa}$ and $\boldsymbol{\gamma}$ are the pixel values of a pixelized convergence map and the observed shear field, respectively.

We can express the linear data model in matrix notation,

\begin{equation} \label{eq:linear_matrix}
\boldsymbol{\gamma} = \mathbf{A} \boldsymbol{\kappa} + \mathbf{n} \ \ ,
\end{equation}

\noindent where the matrix operation $\mathbf{A}$ corresponds to the linear transformation from the ideal (noise-free and full-sky) convergence field to the shear field (equation~\ref{eq:mass_map_operator}). The noise term $\mathbf{n}$ is the vector of noise contributions per pixel (equation~\ref{eq:shape_noise}).

Assume that the average shape noise per pixel on the celestial sphere (e.g. per \healpix{} \cite{GORSKI2005} pixel) is Gaussian distributed, so that the likelihood (dropping $\mathcal{M}$ for brevity) is given by

\begin{equation} \label{eq:likelihood}
p( \boldsymbol{\gamma} | \boldsymbol{\kappa} ) = \frac{1}{\sqrt[]{({ \mathrm{det}  2 \pi}  \mathbf{N})}} \mathrm{exp} \Big[ - \frac{1}{2} ( \boldsymbol{\gamma} - \mathbf{A} \boldsymbol{\kappa} )^\dagger \ \mathbf{N}^{-1} ( \boldsymbol{\gamma} - \mathbf{A} \boldsymbol{\kappa} )  \Big] \  
\end{equation}

\noindent where it is assumed that the noise covariance $\mathbf{N} = \langle \mathbf{n} \mathbf{n}^\dagger \rangle$ is known and that the average noise per pixel is both Gaussian and  uncorrelated (so that $\mathbf{N}$ is diagonal). With this likelihood, the masked (unobserved) pixels have infinite variance.

Under the assumption that the variance per galaxy due to weak lensing is negligible in comparison to the variance due to the intrinsic ellipticity, we can generate noise realizations by rotating the galaxy shapes in the catalogue and thus removing the lensing correlations. This procedure is extremely fast, and allows us to easily construct a Monte Carlo estimate of the noise covariance $\mathbf{N}$.

\subsection{Prior probability distribution}

This work considers four forms for the prior probability distribution $p(\boldsymbol{\kappa} | \mathcal{M})$ that appears in equation~\ref{eq:posterior_optimization}. This prior probability is intrinsic to the method and cannot be `ignored' (in the sense that not including a prior is identical to actively choosing to use a uniform prior).

The various prior probability distributions used in this work correspond to various mass mapping methods, with each prior arising from a different physically motivated constraint. They are:

\begin{enumerate}
\item \noindent Direct Kaiser-Squires inversion. In the absence of smoothing this corresponds to a \textit{maximum a posteriori} estimate with a uniform prior:

\begin{equation} \label{eq:uniform}
p(\boldsymbol{\kappa}) \propto 1 \ \ .
\end{equation}

\noindent Although this is an \textit{improper prior} as it is cannot be normalized, the resulting posterior is nevertheless normalizable. One may set wide bounds for this distribution and in practice these would not impact the final result.

Usually the Kaiser-Squires inversion is followed by a smoothing of small angular scales, where it is expected that noise dominates over signal. 
This corresponds to a lower bound on the prior with respect to angular scale.
\\
\item \noindent E-mode prior (null B-modes). As discussed further in Sec.~\ref{sec:nobmodes}, this prior incorporates our knowledge that weak gravitational lensing produces negligible B-mode contributions. This corresponds to the log-prior

\begin{equation} \label{eq:logbmodeprior}
- \mathrm{log}\  p(\boldsymbol{\kappa}) =   i_{\Imag(\kappa)=0} \ + \  \mathrm{constant} \ \ ,
\end{equation}

\noindent where the indicator function $i_{\Imag(\kappa)=0}$ is discussed in Sec.~\ref{sec:nobmodes}.
\\
\item \noindent Gaussian random field prior, assuming a certain E-mode power spectrum (and with zero B-mode power). The \textit{maximum a posteriori} estimate under such a prior (combined with our Gaussian likelihood) corresponds to a Wiener filter \citep{wiener1949extrapolation,zaroubi_wiener}. The prior distribution

\begin{equation}
p( \boldsymbol{\kappa} ) = \frac{1}{\sqrt[]{({ \mathrm{det}  2 \pi}   \mathbf{S}_\kappa)}} \mathrm{exp} \Big[ - \frac{1}{2} \boldsymbol{\kappa}^\dagger \ \mathbf{S}_\kappa^{-1} \boldsymbol{\kappa}  \Big] \ ,
\end{equation}

\noindent with the power spectrum contributing to the signal covariance matrix $\mathbf{S}_\kappa$, will be discussed in Sec.~\ref{sec:wiener}.
\\
\item \noindent Sparsity-enforced wavelet `halo' prior with null B-modes. In the late Universe it is expected that quasi-spherical halo structures form. A wavelet basis whose elements have this quasi-spherical structure in direct (pixel) space should be a sparse representation of the convergence $\boldsymbol{\kappa}$ signal. This is included in the log-prior distribution

\begin{equation} 
- \mathrm{log}\  p(\boldsymbol{\kappa}) =   \lambda||\boldsymbol{\phi}^\dagger  \boldsymbol{\kappa}||_1  + i_{\Imag(\kappa)=0} \ \ \ ,
\end{equation}

\noindent where the $l_1$ norm of the wavelet transformed convergence $\boldsymbol{\phi}^\dagger  \boldsymbol{\kappa}$ is small when the convergence field contains quasi-spherical halo structures, for a suitable choice of wavelet transform $\boldsymbol{\phi}^\dagger$. Unlike the case of the Gaussian prior, where the lack of B-modes can be included in the power spectrum, here the second term is added to enforce that the signals compatible with the prior contain only E-modes. This is further discussed in Sec.~\ref{sec:sparsity}.

\end{enumerate}

\noindent In the rest of this section we will explain the physical motivation for these choices and show how they are implemented.

\subsection{Kaiser-Squires on the sphere}
\label{sec:KS_method}

In the flat sky limit, for relatively small sky coverage, the $\eth$ operators on the sphere may be approximated using partial derivatives $\partial$ with respect to $\theta$ and $\phi$. In this regime the relationship between shear $\gamma$ and convergence $\kappa$ (equations ~\ref{eq:kappa} and~\ref{eq:gamma}) reduce to

\begin{equation} \label{eq:flat_sky}
\tilde{\gamma} (\mathbf{k}) = \frac{k_1^2 -k_2^2 + 2 i k_1 k_2}{k_1^2 + k_2^2} \tilde{\kappa} (\mathbf{k})  \  \  ,
\end{equation} 

\noindent where $k_1$ and $k_2$ are the components of $\mathbf{k}$, defined in terms of the Fourier transform

\begin{equation} \label{eq:fourier_transform}
 \tilde{\kappa} (\mathbf{k})   = \int_{R^2} \mathrm{d} \boldsymbol{\theta} \ \kappa(\boldsymbol{\theta}) \ \ \mathrm{exp} \big[  i \boldsymbol{\theta} \cdot \mathbf{k} \big] \  \  ,
\end{equation} 

\noindent where $\boldsymbol{\theta} $ has components $\theta$ and $\varphi$. The well-known Kaiser-Squires (KS) method estimates the convergence by directly inverting equation~\ref{eq:flat_sky}.

For the DES Y3 sky coverage, the flat sky approximation cannot be used without introducing substantial errors \citep{Wallis2017}, so as in the Y1 mass map analysis \citep{Chang2018} we require a curved-sky treatment. KS on the sphere corresponds to a decomposition of the spin-2 field $\gamma$ into a curl-free E-mode component and a divergence-free B-mode component, as described in Sec.~\ref{sec:theory}.

With these components $\hat{\gamma}_{\mathrm{E}, \ell m}$ and $\hat{\gamma}_{\mathrm{B}, \ell m}$ we use equation~\ref{eq:mass_map_operator} to recover $\hat{\kappa}_{\mathrm{E}, \ell m}$ and $\hat{\kappa}_{\mathrm{B}, \ell m}$, which transform as scalars using a spin-0 spherical harmonic transform to recover $\kappa(\theta, \varphi) = \kappa_\mathrm{E}(\theta, \varphi)  + i \ \kappa_\mathrm{B}(\theta, \varphi)$.

The spherical harmonic operations described above are entirely analogous to CMB linear polarization, where the Q and U Stokes parameters correspond to the $\gamma_1$ and $\gamma_2$ components. As such, all spherical harmonic transformations use either the scalar or `polarization' transforms of \healpix{}~\citep{GORSKI2005}. All maps presented in this work use $\nside{} = 1024$ and all relevant spherical harmonic transforms use $\ell_{max}=2048$.

As with flat-sky KS, this generalization of KS to the celestial sphere corresponds to an inverse of the linear operation $\mathbf{A}$ in equation~\ref{eq:linear_matrix} and, as such, corresponds to a maximum likelihood estimate (c.f. equation~\ref{eq:likelihood}) of the convergence field $\kappa$. Direct KS inversion therefore corresponds to a \textit{maximum a posteriori} estimate with a uniform prior $p(\boldsymbol{\kappa}) \propto 1$.

{  Even with this Bayesian \textit{maximum a posteriori} interpretation, the KS reconstruction method has the advantage of simplicity: the transformation is linear if B-modes are included (which can be a useful mathematical property) and the method is computationally straightforward.}

As is standard practice the KS inversion is followed by a smoothing of small angular scales, corresponding to a lower bound on the prior with respect to angular scale. We treat the choice of the angular smoothing parameter as a free parameter, the effects of which we investigate using simulated data (Sec.~\ref{sec:sims_maps}).

\subsection{Null B-mode prior} \label{sec:nobmodes}

We can decompose a convergence map into a real E-mode and imaginary B-mode component

\begin{equation}
\kappa = \kappa_\mathrm{E} + i \ \kappa_\mathrm{B} \ \ ,
\end{equation}

\noindent where the shear representation of the E-mode $\kappa_\mathrm{E}$ is curl-free and the B-mode $\kappa_\mathrm{B}$ is divergence-free.

The Born-approximation weak lensing derivation (see Sec.~\ref{sec:theory}) makes it clear that weak gravitational lensing generates no B-mode components. Higher order contributions can contribute to non-zero B-modes (e.g. \citealt{Krause2010}), although these effects are generally much smaller than the leading E-mode contribution. Additionally, intrinsic alignments of galaxies can induce non-zero B-mode contributions \citep{Blazek2017,Samuroff2019}, although intrinsic alignment effects are not included in this map reconstruction analysis. We also note that systematic effects, such as shear measurement systematic errors of point-spread-function residuals, can also generate spurious B-modes (e.g. \citealt{Asgari2019}), but no significant B-modes have been measured in the DES Y3 shear catalogue \citep*{y3-shapecatalog}.

The standard KS reconstruction generates spurious B-modes due to shape noise and masks. It is therefore well-motivated to have a prior probability distribution for convergence $\kappa$ that gives no probability to $\kappa_\mathrm{B}$ and the KS uniform prior to $\kappa_\mathrm{E}$ only, giving the following log-prior

\begin{equation}
- \mathrm{log}\  p(\boldsymbol{\kappa}) =   i_{\Imag(\kappa)=0} \ + \  \mathrm{constant} \ \ ,
\end{equation}

\noindent where the indicator function of a set $\mathcal{C}$ is defined as

\begin{equation} \label{eq:indicator}
i_{\mathcal{C}}(x) = \begin{cases}
	0 & \mbox{if } x \in \mathcal{C}\\
	+ \infty & \mbox{otherwise}  \ \ \  ,
\end{cases}
\end{equation}

\noindent which in our case gives zero prior probability to convergence $\kappa$ maps with an imaginary component (corresponding to B-modes). The \textit{maximum a posteriori} estimate with this prior and Gaussian likelihood is given by the following optimization problem:

\begin{equation}
\hat{\boldsymbol{\kappa}} = \underset{\kappa}{\rm arg \ min} \  ( \boldsymbol{\gamma} - \mathbf{A} \boldsymbol{\kappa} )^\dagger \ \mathbf{N}^{-1} ( \boldsymbol{\gamma} - \mathbf{A} \boldsymbol{\kappa} ) +  i_{\Imag(\kappa)=0}  \ .
\end{equation}

\noindent This formulation allows us to maximize the log posterior (equation~\ref{eq:posterior_optimization}) using Forward-Backward Splitting (\citealt{combettes2005signal}), with a proximity operator corresponding to an orthogonal projector onto the set $\mathcal{C}$. This is implemented with the following iterative method

\begin{equation}
\boldsymbol{\kappa}^{\small (n+1)} = {\mathrm{Re}} \Big[ \boldsymbol{\kappa}^{\small (n)} + \mu \mathbf{A}^\dagger \mathbf{N}^{-1} \big( \boldsymbol{\gamma} - \mathbf{A}  \boldsymbol{\kappa}^{\small (n)} \big)  \Big] \ \ ,
\end{equation}

\noindent where $\mu$ controls the gradient steps and is free to be chosen {  within certain broad conditions (see~\citealt{combettes2005signal}), which allows us to represent the iterative method as

\begin{equation}
\boldsymbol{\kappa}^{\small (n+1)} = {\mathrm{Re}} \Big[ \boldsymbol{\kappa}^{\small (n)} + \mu' \mathbf{A}^\dagger \big[ \mathbf{n}_g  \odot \big( \boldsymbol{\gamma} - \mathbf{A}  \boldsymbol{\kappa}^{\small (n)} \big)  \big] \Big] \ \ ,
\end{equation}

\noindent where $\odot$ is an element-wise (Hadamard) product. Here we have absorbed the amplitude of the noise variance into $\mu'$ leaving just a vector of number of galaxies per pixel $\mathbf{n}_g$ with galaxy weights according to Sec.~\ref{sec:data_sims}. In practice, the second term can be numerically unstable due to the forward and backward transforms  ($\mathbf{A}$, $\mathbf{A}^\dagger$) on the \healpix{} sphere, becoming increasingly problematic for low signal-to-noise data, which necessitates some regularization of the gradient update steps. As with KS, we ultimately smooth small scales of the reconstructed map, and we therefore initialize $\boldsymbol{\kappa}^{(0)}$ with the smoothed KS reconstruction and include the smoothing operation after each gradient update step which also serves as a regularizer in the gradient descent. This also implies that the final map would be slightly smoother than if it had been smoothed only at the end of the iterative procedure.
}

Although the motivation and the algorithm are somewhat different, this method is inspired by and gives a similar outcome to that shown in~\cite{MawdsleyFF_2019arXiv190512682M}. The algorithm described here is also similar to the \textit{GKS} special case of the MCALens method for flat-sky mass mapping as described in the appendices of~\cite{starck_mcalens}.

\subsection{Gaussian prior (Wiener filter)}
\label{sec:wiener}

This prior is that of a Gaussian random field, which is applicable for the density field on large scales at late times, 

\begin{equation}
p( \boldsymbol{\kappa} | \mathbf{S}_\kappa ) = \frac{1}{\sqrt[]{({ \mathrm{det}  2 \pi}  \mathbf{S}_\kappa)}} \mathrm{exp} \Big[ - \frac{1}{2} \boldsymbol{\kappa}^\dagger \ \mathbf{S}_\kappa^{-1} \boldsymbol{\kappa}  \Big] \ .
\end{equation}

The \textit{maximum a posteriori} estimate with this prior and Gaussian likelihood is given by the following optimization problem:

\begin{equation}
\hat{\boldsymbol{\kappa}} = \underset{\kappa}{\rm arg \ min} \  ( \boldsymbol{\gamma} - \mathbf{A} \boldsymbol{\kappa} )^\dagger \ \mathbf{N}^{-1} ( \boldsymbol{\gamma} - \mathbf{A} \boldsymbol{\kappa} ) + \boldsymbol{\kappa}^\dagger \ \mathbf{S}_\kappa^{-1} \boldsymbol{\kappa}  \ .
\end{equation}

\noindent The solution to this problem is the Wiener filter:

\begin{equation} \label{eq:wiener}
\begin{split}
\hat{\boldsymbol{\kappa}}_W &= \mathbf{W} \boldsymbol{\gamma} \\
\mathbf{W} &= \mathbf{S}_\kappa \mathbf{A}^\dagger \big[ \mathbf{A} \mathbf{S}_\kappa \mathbf{A}^\dagger + \mathbf{N} \big]^{-1} \ .
\end{split}
\end{equation}

\noindent Here $\mathbf{S}_\kappa$ and $\mathbf{N}$ are the signal and noise covariance matrices respectively, which are $\langle \boldsymbol{\kappa} \boldsymbol{\kappa}^\dagger \rangle $ and $\langle \mathbf{n} \mathbf{n}^\dagger \rangle$ for this problem.  

Direct evaluation of the matrix $\mathbf{W}$, which has at least $10^{12}$ elements and is sparse in neither pixel space nor harmonic space, would be extremely computationally expensive. We therefore make use of a class of methods that use additional {\textit{messenger fields}} (introduced by~\citealt{elsner_wiener}) to iteratively transform between pixel space, where $\mathbf{N}$ is diagonal, and harmonic space, where $\mathbf{S}_\kappa$ is diagonal. Such methods have seen widespread use in cosmology where the signal covariance is often sparse due to the statistical isotropy of the underlying signal ~\citep{jasche2015matrix, alsing2017cosmological, jeffrey_heavens_fortio}.

For a Wiener filter messenger field implementation on the sphere we the {\sc Dante}\footnote{\url{https://github.com/doogesh/dante}} package~\citep{dante}, which uses an optimized novel messenger field implementation to perform Wiener filtering on the sphere for spin-2 fields. {  We test convergence by doubling the {\sc Dante} precision (with precision parameter from $10^{-5}$ to $5\times10^{-6}$), which effectively corresponds to increasing the number of iterations, and showing a negligible MSE change of $3\times 10^{-5}$ per cent with simulated data.

The signal covariance matrix in harmonic space is diagonal, with elements given by an assumed fiducial power spectrum. Our fiducial E-mode power spectrum is taken as the power spectrum of the convergence truth map from the simulated data (see Sec.~\ref{sec:data_sims}) which was corrected for the mask using the \texttt{NaMaster}\footnote{\url{https://github.com/LSSTDESC/NaMaster}} pseudo-$C_\ell$ estimation code~\citep{Alonso_2019}.}

We explicitly provide a B-mode power spectrum set to zero, thus simultaneously achieving the null B-mode prior equivalent to Sec.~\ref{sec:nobmodes}.

\subsection{Sparsity prior} \label{sec:sparsity}

The optimization problem solved by the \glimpse{} algorithm using a sparsity prior is

\begin{equation}
\hat{\boldsymbol{\kappa}} = \underset{\kappa}{\rm arg \ min} \  ( \boldsymbol{\gamma} - \mathbf{A} \boldsymbol{\kappa} )^\dagger \ \mathbf{N}^{-1} ( \boldsymbol{\gamma} - \mathbf{A} \boldsymbol{\kappa} ) + \lambda || \mathbf{\omega} \mathbf{\Phi}^\dagger \boldsymbol{\kappa} ||_1  +   i_{\rm{Im}(\boldsymbol{\kappa}) = 0}  \  ,
\end{equation}

\noindent where $\mathbf{\omega}$ is a diagonal matrix of weights, and $\mathbf{\Phi}^\dagger$ is the inverse wavelet transform. The indicator function $i_{\rm{Im}(\cdot) = 0}$ in the final term imposes realness on the reconstruction (null B-modes). The use of nonuniform discrete Fourier transform (NDFT) allows the first term to perform a forward-fitted Kaiser-Squires-like step without binning the shear data, allowing the smaller scales to be retained in the reconstruction. The full algorithm, including the calculation of the weights, is described in Sec. 3.2 in \cite{lanusse_2016}.

\glimpse{} operates on a small patch of the sky, which it treats as flat. Input shear data is transferred (projected) from the celestial sphere to the tangent plane (i.e. the plane tangent to the sphere at the patch centre); the `shear to convergence' calculation is done on the tangent plane (where the flatness simplifies the analysis); the results (which are reported at a lattice of points - call this an `output lattice') are then mapped back to the sphere. The mapping between sphere and tangent plane is the orthographic projection.

To analyse the large DES footprint we run \glimpse{} on multiple (overlapping) small patches and paste the results together. We set each of our patches to be 256 square degrees (a compromise: larger would stress the flat-sky approximation while smaller would suppress large-scale modes). The density of such patches is one per 13 square degrees. The output lattices were set to have $330 \times 330$ points. Each pixel in our draft convergence map (\healpix{} $\nside{} = 2048$) is obtained from a weighted average of the convergences at all the output lattice points, from all the patches, that happen to fall in that pixel. The weights are chosen to be unity in the centre of each patch but to fall away to zero (sharply but smoothly) away from the central one-ninth of each output patch. As a last step the output convergence map is downsampled to a $\nside{} = 1024$.

An alternative to this patching strategy would be to implement wavelets on the sphere. The sparsity-based statistical model described by~\cite{price_sphere} demonstrate such a strategy, with the added benefit of sampling the posterior distribution (not just maximization), though uses wavelets on the sphere that have infinite support in pixel space.

The choice of wavelet transformation (sometime called a `dictionary') depends on the structures contained in the signal. Theory predicts the formation of quasi-spherical haloes of bound matter. It is standard practice to represent the spatial distribution of matter in haloes with spherically symmetric Navarro-Frenk-White~\citep{nfw} or Singular Isothermal Sphere profiles. The starlet, Coefficients of Isotropic Undecimated Wavelets~\citep{starck_sparsity}, in two dimensions are well suited to represent the observed convergence of a dark matter halo. The wavelet transform used in the \glimpse{} algorithm is the starlet~\citep{starck2007undecimated}, which can represent positive, isotropic objects well. This prior in the starlet basis represents a physical model that the matter field is a superposition of spherically symmetric dark matter haloes.

The full \glimpse{} algorithm is described in detail in \cite{lanusse_2016}.

{ 
\subsection{Properties of inferred maps}\label{sec:prop}
As described above, each of our maps is a \textit{maximum a posteriori} estimate given the observed data; that is, each is the most probable map for the data given one of our assumed models. All mapping methods take into account the same noise covariance matrix (characterising the noise amplitude and distribution across the observed area); differences between the maps arise from the different assumptions about the prior probability distribution for the underlying convergence $\boldsymbol{\kappa}$. 

Although the map (in practice this is a set of pixel values) is the most probable map, a given statistic of the map will not necessarily correspond to the most probable statistic. For example, if the convergence $\kappa$ field is indeed Gaussian, we can see that the resulting most probable map is the Wiener filter map. The two-point statistics (e.g. power spectrum) of the Wiener filtered map will comprise terms such as $\langle \hat{\boldsymbol{\kappa}} \hat{\boldsymbol{\kappa}}^\dagger \rangle = \langle \mathbf{W} \boldsymbol{\gamma}   \boldsymbol{\gamma}^\dagger \boldsymbol{\mathbf{W}}^\dagger \rangle $. If the signal-to-noise ratio is not infinite (i.e. $\mathbf{S} + \mathbf{N} \neq \mathbf{S}$), equation~\ref{eq:wiener} for $\mathbf{W}$ shows that the two-point statistics of the Wiener filtered map $\langle \hat{\boldsymbol{\kappa}} \hat{\boldsymbol{\kappa}}^\dagger \rangle$ will have lower amplitude than those of the truth $\langle {\boldsymbol{\kappa}} {\boldsymbol{\kappa}}^\dagger \rangle$.

This is no contradiction: the pixel values forming their most probable combination $\hat{\boldsymbol{\kappa}}$ maximize $p(\boldsymbol{\kappa} | \boldsymbol{\gamma})$, but would not maximize a transformed probability $p(\boldsymbol{\kappa}^2 | \boldsymbol{\gamma})$. For most summary statistics, the map cannot simultaneously be the most probable map and be trivially used to derive the most probable summary statistic. If we evaluated the full posterior $p(\boldsymbol{\kappa}|\boldsymbol{\gamma})$ rather than evaluating a \textit{maximum a posteriori} point-estimate, we could transform the probability density to further evaluate functions of the map (e.g. spectra, correlation functions, moments).

If we wished to jointly estimate the map and a given statistic $\mu$ used in the map-making process (e.g. $C_\ell$ for Wiener filtering or $\lambda$ for the sparsity prior) we could instead form the joint posterior $p(\boldsymbol{\kappa}, \mu | \boldsymbol{\gamma})$ and jointly estimate $\mu$. It has been demonstrated that under certain assumptions one can indeed jointly sample the lensing map and the unknown power spectrum \citep{wandelt_sampling, alsing2017cosmological} or the unknown $\lambda$ parameter (e.g. \citealt{higson, price_maps}) if this is desired. In this work we evaluate a point-estimate that maximizes $p(\boldsymbol{\kappa} | \boldsymbol{\gamma})$ and, as we do not aim to evaluate the full posterior, we fix $C_\ell$ (even doubling the amplitude leads to sub-$5$-percent change in mean-square-error for the point estimate) and tune $\lambda$ using simulated data (Sec.~\ref{sec:sims_maps}).

For inference using map-based statistics, the theoretical predictions can be simply adjusted for the given map reconstruction. In a forward-modelling framework (as used by many higher-order statistics), the predictions are measured from mock maps and the same operations are applied consistently to the mock data and to the observed data.
}

\section{Data and simulations}
\begin{figure}
\includegraphics[width=0.43\textwidth]{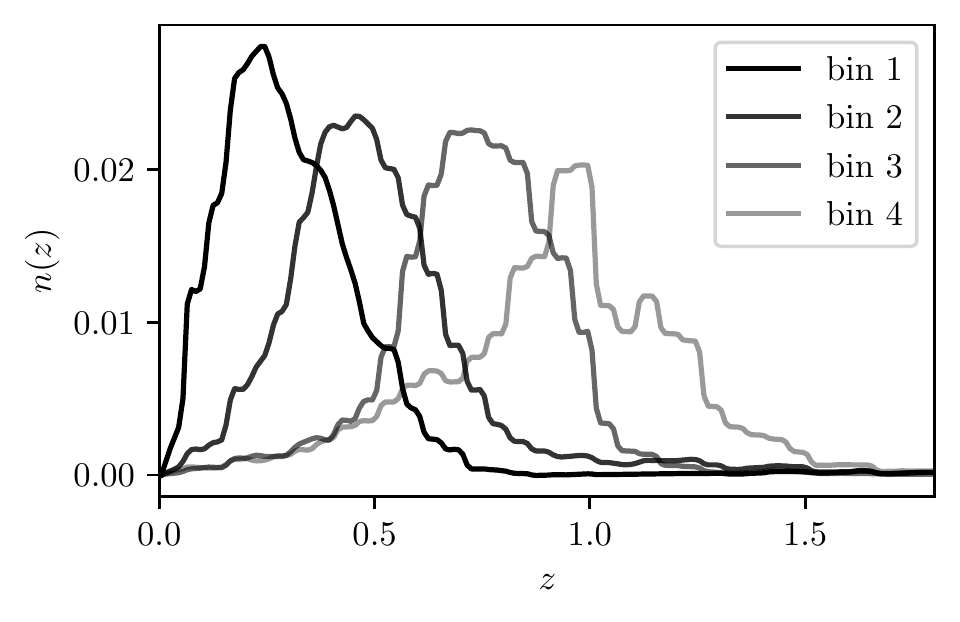}
\caption{Redshift distributions as estimated in data for the four DES Y3 tomographic bins \citep{y3-sompz}.}
\label{fig:Nz_sims_data}
\end{figure}

\label{sec:data_sims}
In this paper we used data products from the first three years (Y3) of the Dark Energy Survey \citep[DES,][]{DES_review,DES_DR1}, and mock galaxy catalogues that were tailored to match the data. DES is a five-year survey that covers $\sim 5000~\mathrm{\deg}^2$ of the South Galactic Cap. Mounted on the Cerro Tololo Inter-American Observatory (CTIO) four metre Blanco telescope in Chile, the $570$~megapixel Dark Energy Camera \citep[DECam,][]{Flaugher2015} images the field in $grizY$ filters. The raw images were processed by the DES Data Management (DESDM) team  \citep{Sevilla2011,Morganson2018,DES_DR1}. For the full details of the data, we refer the readers to \cite{y3-gold} and \cite*{y3-shapecatalog}.

\subsection{The DES Y3 shear catalogue}
The DES Y3 shear catalogue, described in detail in \cite*{y3-shapecatalog}, builds upon the Y3 Gold catalogue \citep{y3-gold}. It is created using the \texttt{METACALIBRATION} algorithm \citep{HuffMcal2017, SheldonMcal2017}, which infers the galaxy ellipticities starting from noisy images of the detected objects in the \textit{r, i, z} bands. The \texttt{METACALIBRATION} algorithm was used previously in the DES Y1 analysis \citep*{ZuntzY1}. \texttt{METACALIBRATION} provides an estimate of the shear field, and it relies on a self calibration framework using the data itself to correct for the response of the estimator to shear as well as for selection effects. Objects are included in the catalogue if they pass a number of selection cuts designed to reduce potential systematic biases \citep*{y3-shapecatalog}. Inverse variance weights are assigned to galaxies. The final DES Y3 shear catalogue has 100,204,026 objects, with a weighted $n_{\rm eff}=5.59$~galaxies~arcmin$^{-2}$.

Despite the \texttt{METACALIBRATION} response self-correcting for most of the multiplicative bias, it is known that for the DES Y3 shear catalogue there is an additional multiplicative bias of approximately $2$ or $3$ per cent \citep{y3-imagesims}. This factor arises partly from a shear-redshift-dependent detection bias due to blending of galaxy images, for which the \texttt{METACALIBRATION} implementation adopted in DES Y3 is unable to account \citep{SheldonMetadetect2019}. This multiplicative factor is left uncalibrated but is marginalised over in the main cosmological analysis. In \cite*{y3-shapecatalog} the shear catalogue has also been tested for additive biases (e.g. due to point-spread-function residuals). In particular, the catalogue is characterized  by a non-zero mean shear whose origin is unknown and which is subtracted at the catalogue level before performing any analysis. 

A two-stage blinding procedure was used in the DES Y3 analysis to mitigate confirmation bias. The first level of this procedure blinded the shear catalogue by means of a multiplicative factor, in a fashion similar to what has been adopted in the Y1 analysis \citep*{ZuntzY1}. The second level of blinding \citep{Muir2019} was applied to the summary statistics under examination (e.g. cosmic shear, galaxy-galaxy lensing, galaxy-galaxy clustering).  {Since in this work we do not directly measure any summary statistics from the data maps, only the first level of blinding has been considered. All the systematic tests on the maps obtained from the data have been performed first using the blinded catalogue, and then repeated after unblinding.  }

 {The shear catalogue is further divided into four tomographic bins; redshift distribution estimates (Fig. \ref{fig:Nz_sims_data}) for each of the tomographic bins are provided by the SOMPZ method \citep*{y3-sompz}, further informed by clustering (WZ) constraints \citep*{y3-sourcewz}. The $n(z)$ are also tweaked to take into account the redshift-dependent effect of blending \citep{y3-imagesims}. When running the cosmological analysis, constraints on the $n(z)$ are further improved by shear-ratio constraints \citep{y3-shearratio}. The tomographic bins are selected so as to have {roughly} equal number density.}

 {The catalogue is then used to create shear maps (i.e. pixelized maps for the two components of the shear field). The maps are constructed using a \healpix{} pixelization \citep{GORSKI2005} with $\nside{} = 1024$ (corresponding to a pixel size of $3.44~\mathrm{arcmin}$). The estimated value of the shear field in the map pixels is given by:}
\begin{equation}
\label{eq:pixelvalue}
\gamma_{\rm obs}^{\nu} = \frac{\sum_{j=1}^{n}\epsilon_j^{\nu}w_{j}}{\bar{R}\sum_{j=1}^{n}w_j}, \,\, \nu=1,2,
\end{equation}
 {where $\nu$ refers to the two shear field components, $n$ is the total number of galaxies in the sample, $w_j$ is the per-galaxy inverse variance weight, and $\bar{R}$ is the average \texttt{METACALIBRATION} response of the sample. Eq.~\ref{eq:pixelvalue} is used to create shear field maps for the full catalogue as well as for the four tomographic bins. As mentioned earlier, the multiplicative shear bias is left uncalibrated when creating the shear maps. Any non-zero mean shear is subtracted from the catalogue before creating the maps.}

\subsection{Simulated mock galaxy catalogue}
 {
To build our simulated galaxy catalogue, we use a single realization of the 108 available \cite{Takahashi2017} simulations. These are a set of full-sky 
lensing convergence and shear maps obtained for a range of redshifts between $z = 0.05$ and $5.3$ at intervals of $150~ h^{-1}~\mathrm{Mpc}$ comoving distance.

Initial conditions were generated using the 2LPTIC code \citep{Crocce2006} and the N-body simulation used L-GADGET2 \citep{springel2005} with cosmological parameters consistent with WMAP 9 year results \citep{Hinshaw2013}: $\Omega_{\rm m} = 0.279$, $\sigma_8 = 0.82$, $\Omega_{\rm b} = 0.046$, $n_{\rm s} = 0.97 $, $h = 0.7$. The simulations begin with 14 boxes with side lengths $L = 450, 900, 1350, ..., 6300~h^{-1}~\mathrm{Mpc}$ in steps of 450~$h^{-1}~\mathrm{Mpc}$, with six independent copies at each box size and $2048^3$ particles per box. Snapshots are taken at the redshift corresponding to the lens planes at intervals of $150~h^{-1}~\mathrm{Mpc}$ comoving distance. The average matter power spectra of the simulations agree with the revised \texttt{HALOFIT} \citep{Takahashi2012} predictions within $5$ per cent for $k < 1~h~\mathrm{Mpc}^{-1}$ at $z < 1$, for $k < 0.8~h~\mathrm{Mpc}^{-1}$ at $z < 3$, and for $k < 0.5~h~ \mathrm{Mpc}^{-1}$ at $z < 7$. A multiple plane ray-tracing algorithm (\texttt{GRayTrix}, \citealt{Hamana2015}) is used to estimate the values of the shear and convergence fields for the simulation snapshots. Shear and convergence field maps are provided in the form of \healpix{} maps with resolution $\nside{} = 4096$. 

We use the convergence and shear maps at different redshifts to generate a simulated DES Y3 shape catalogue, using the following procedure. First, we generate convergence and shear field \healpix{} maps for the four DES Y3 tomographic bins (and for the full catalogue as well) by stacking the shear and convergence snapshots, properly weighted by the fiducial DES Y3 redshift distributions of the bins. Simulated galaxies are then randomly drawn within the DES Y3 footprint according to the DES Y3 number density. Each simulated galaxy is assigned a shear and convergence value depending on its position (i.e. by looking at the value of that particular pixel of the convergence and shear maps into which they fall). To assign realistic shape noise and weights to the simulated galaxies, we make use of the fiducial DES Y3 shape catalogue. In particular, we randomly rotate the ellipticity of each galaxy in the data such that it can be used as intrinsic ellipticity. This intrinsic ellipticity is added to a random galaxy of the simulated catalogue, using the shear addition formula (e.g. \citealt{Seitz1997}). We also assign to the simulated galaxy the inverse variance weight from the same real galaxy we used to obtain the intrinsic ellipticity. Following this procedure, we obtain a simulated DES Y3 catalogue, with the same number density, shape noise and weights of the catalogue in data. Finally, following Eq. \ref{eq:pixelvalue}, we use the simulated catalogue to create a $\nside{} = 1024$ `true' convergence map, which will be used as comparison in all the tests on simulations.}

\section{Simulation tests}
\label{sec:sims_maps}

\begin{figure*}
    \includegraphics[width=0.49\textwidth]{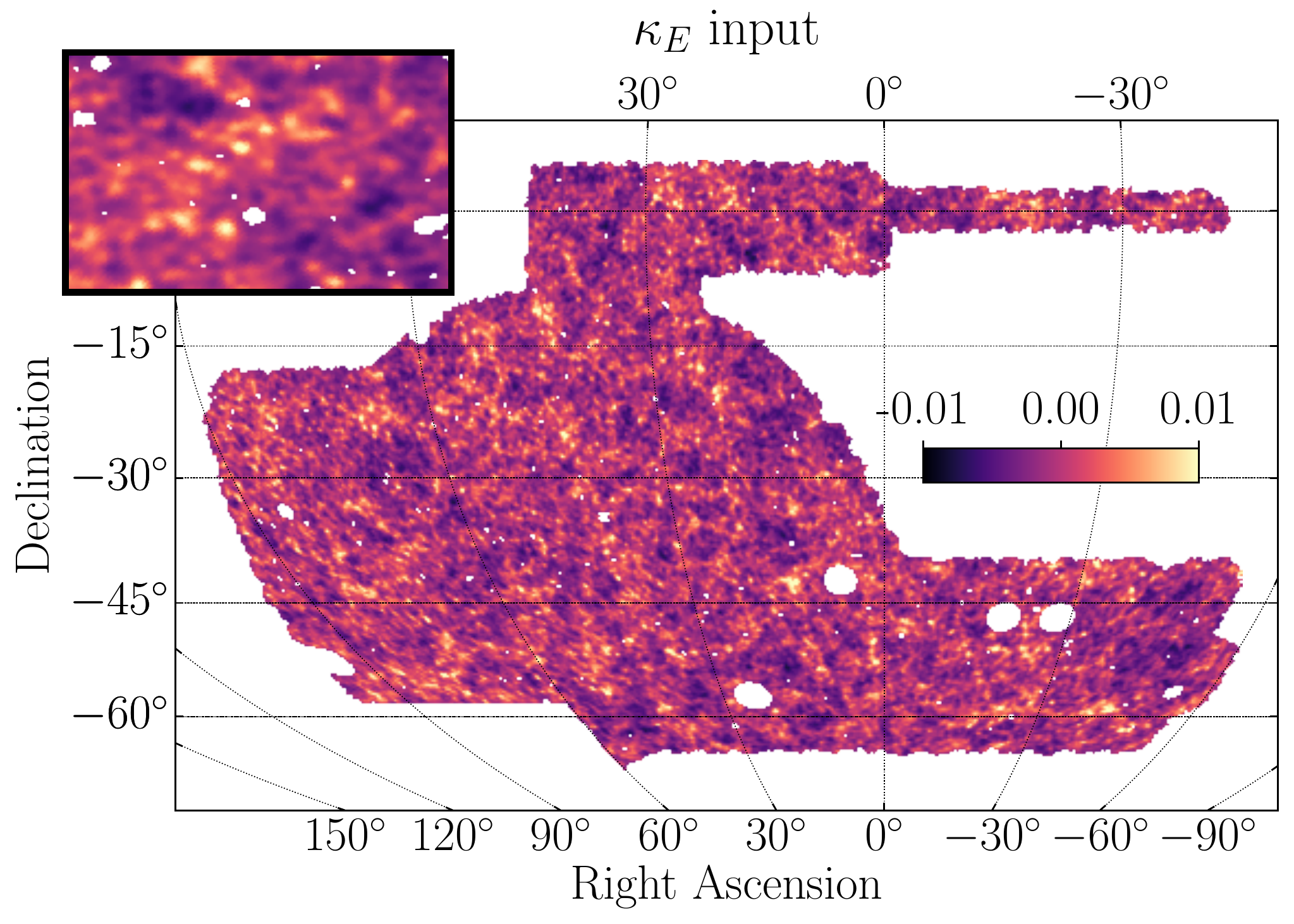}
    \includegraphics[width=0.49\textwidth]{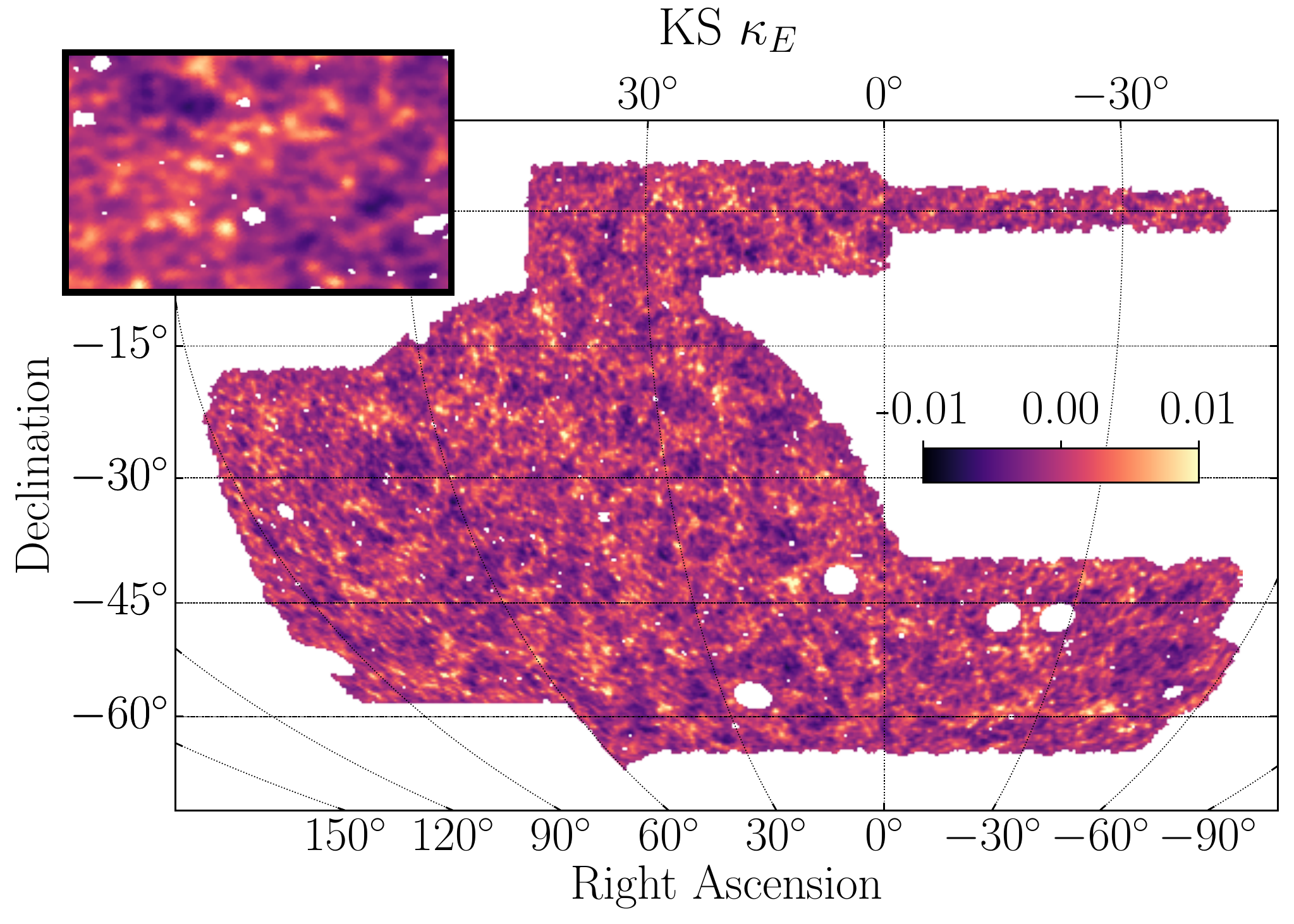}
    \includegraphics[width=0.49\textwidth]{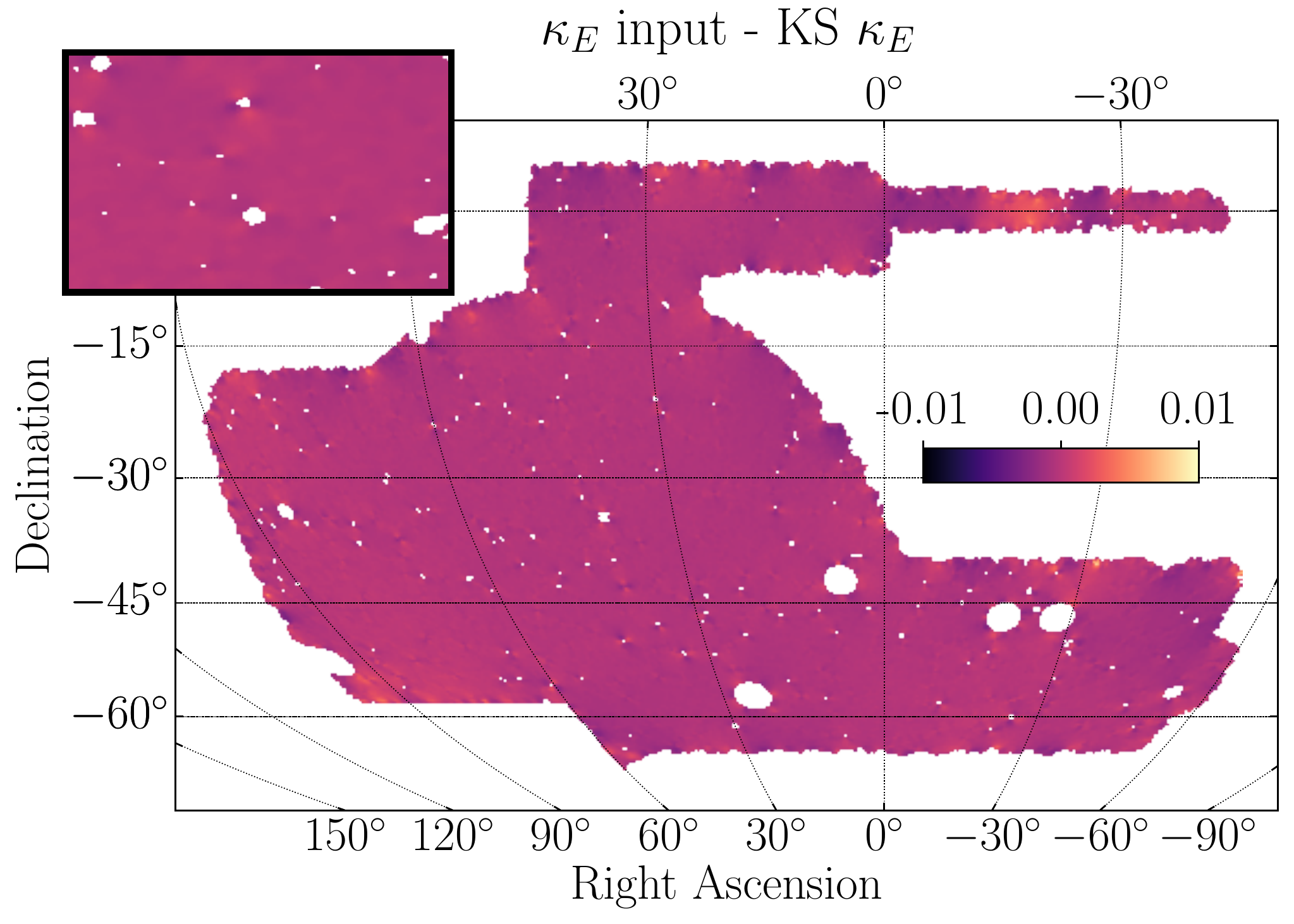}
    \includegraphics[width=0.49\textwidth]{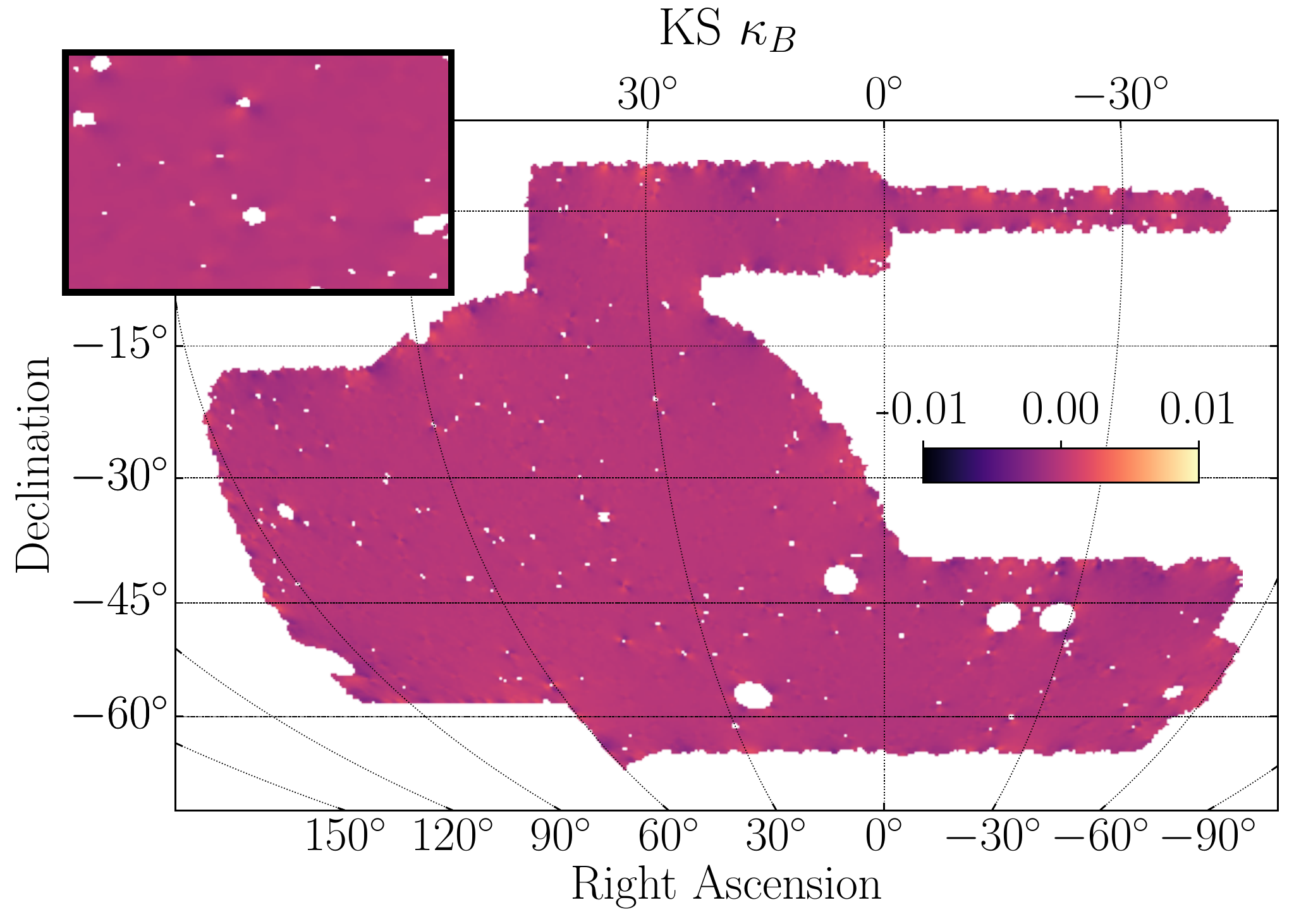} 
    \caption{Simulated noise-free DES Y3 weak lensing mass maps. \textit{Top left panel}: the original input convergence field map. \textit{Top right panel}: the convergence field map (E-mode) obtained using the spherical KS algorithm from a noiseless realization of the shear field. \textit{Bottom left panel}: residual map of the input convergence field and the KS map. \textit{Bottom right panel}: KS B-mode map. Maps have been smoothed at $10~\mathrm{arcmin}$ for visualization purposes. \textit{Inset}:  {\small{RA}}$_\mathrm{centre}$, Dec$_\mathrm{centre} = 70^{\circ}, -40^{\circ}$; $\Delta$RA, $\Delta$Dec = 15$^{\circ}$, 10$^{\circ}$.}
    \label{fig:MM_y3_1}
\end{figure*}
\begin{figure}
\begin{center}
\includegraphics[width=0.48 \textwidth]{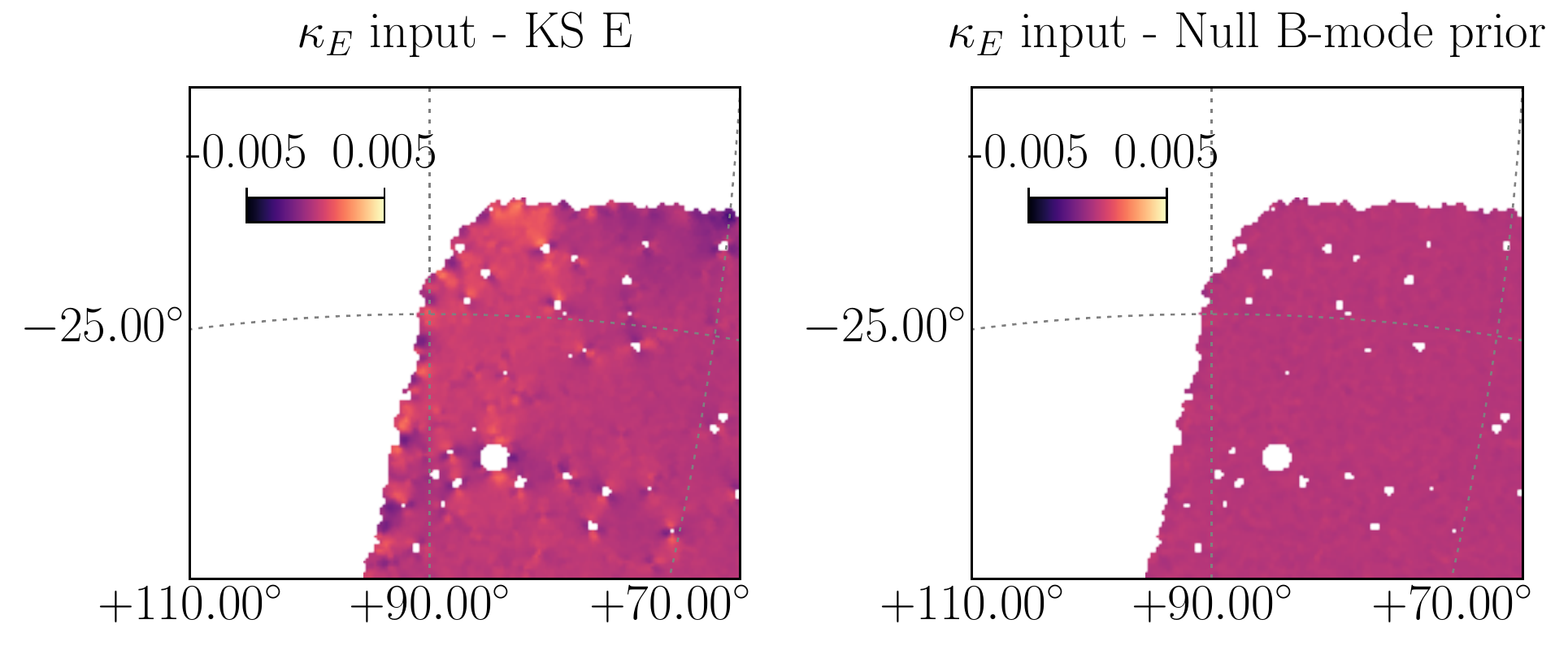}
\end{center}
\caption{Zoomed-in version of the residual maps for the KS (left) and null B-mode prior methods (right). The maps have been zoomed close to the edge of the footprint. The null B-mode prior method is characterized  by a lower amplitude of the residual map, owing to a better handling of the mask effects.}
\label{fig:residuals_zoomed}
\end{figure}
\begin{figure}
\centering
\includegraphics[width=0.4 \textwidth]{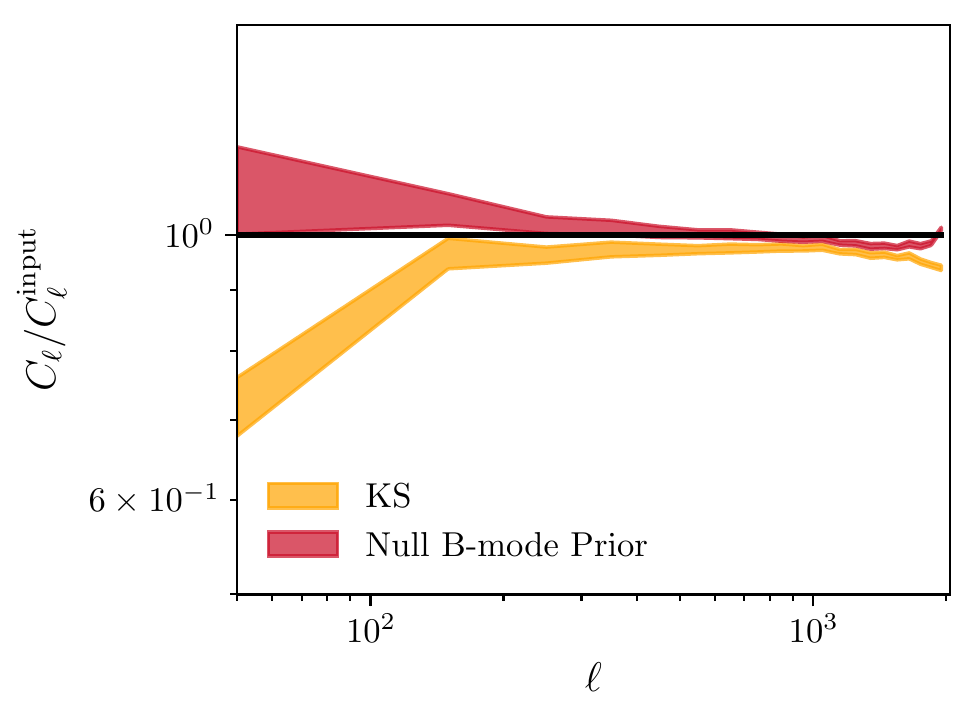}
\vspace{-0.1in}
\caption{Power spectrum of the reconstructed maps, for the KS and the null B-mode prior methods, obtained from a noiseless realization of the shear field. No smoothing has been applied to the recovered maps. We compare here with the power spectra of the input convergence field.}
\label{fig:Cl_2}
\end{figure}
\begin{figure*}
    \includegraphics[width=0.49\textwidth]{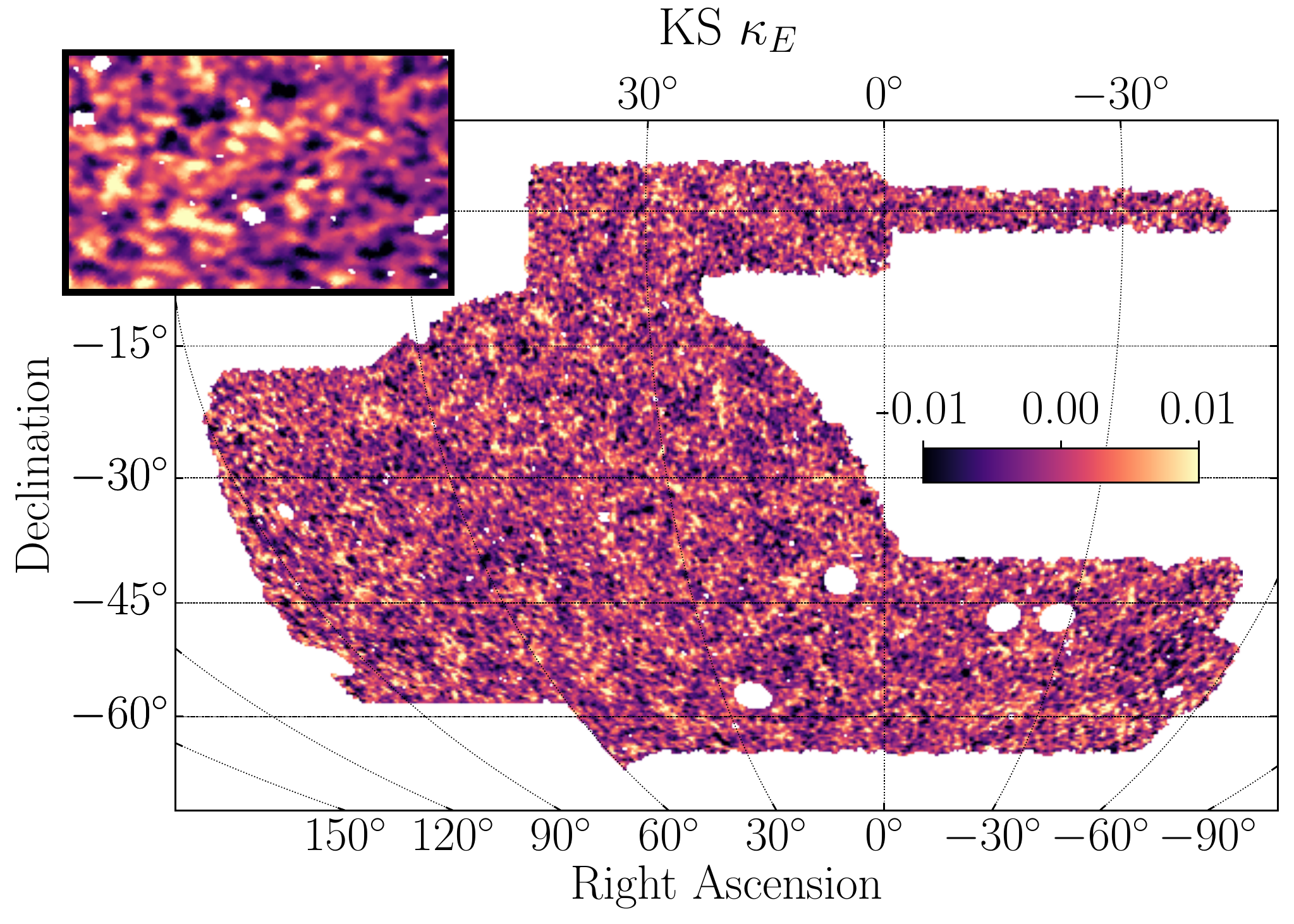}
    \includegraphics[width=0.49\textwidth]{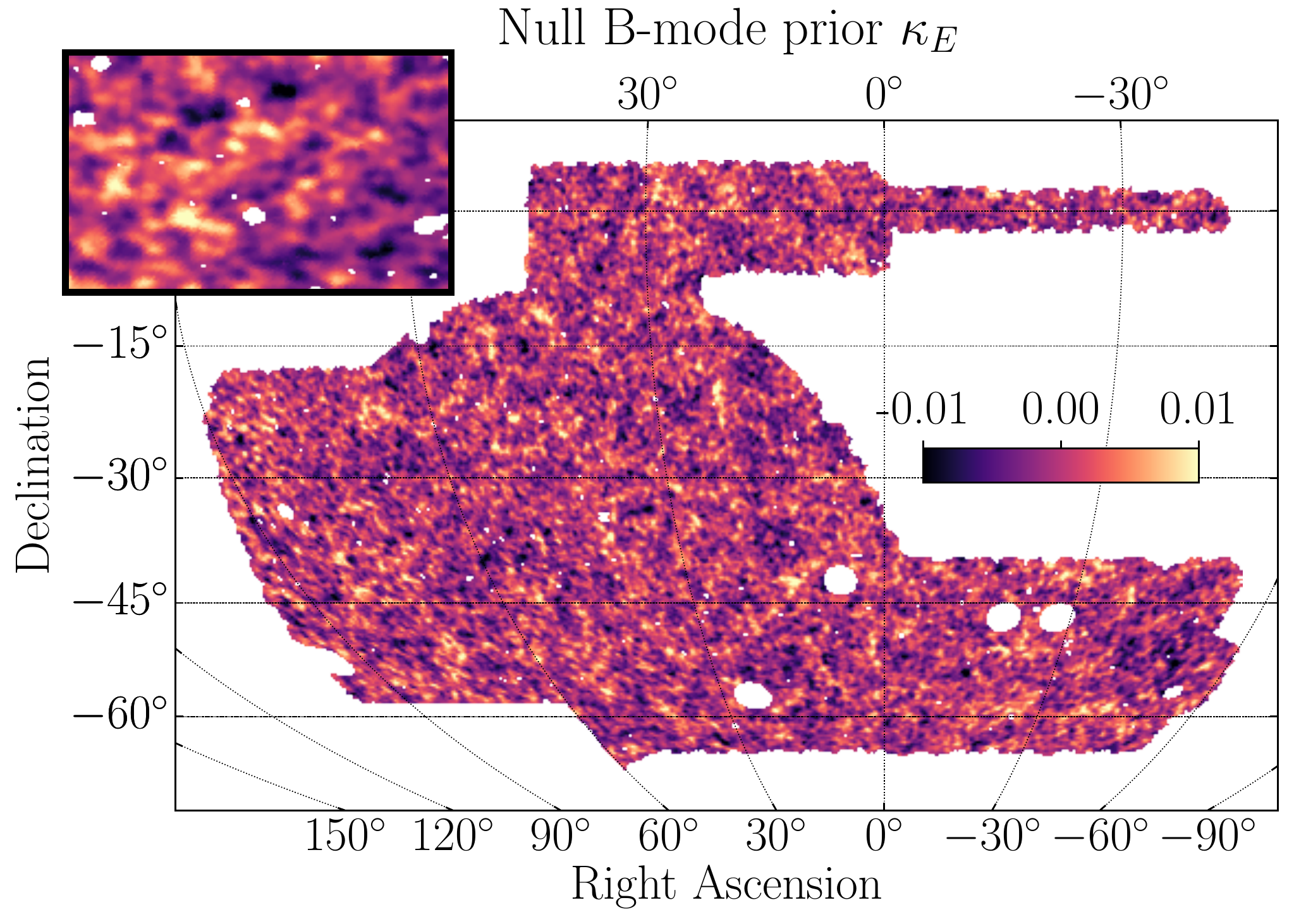} \\
    \includegraphics[width=0.49\textwidth]{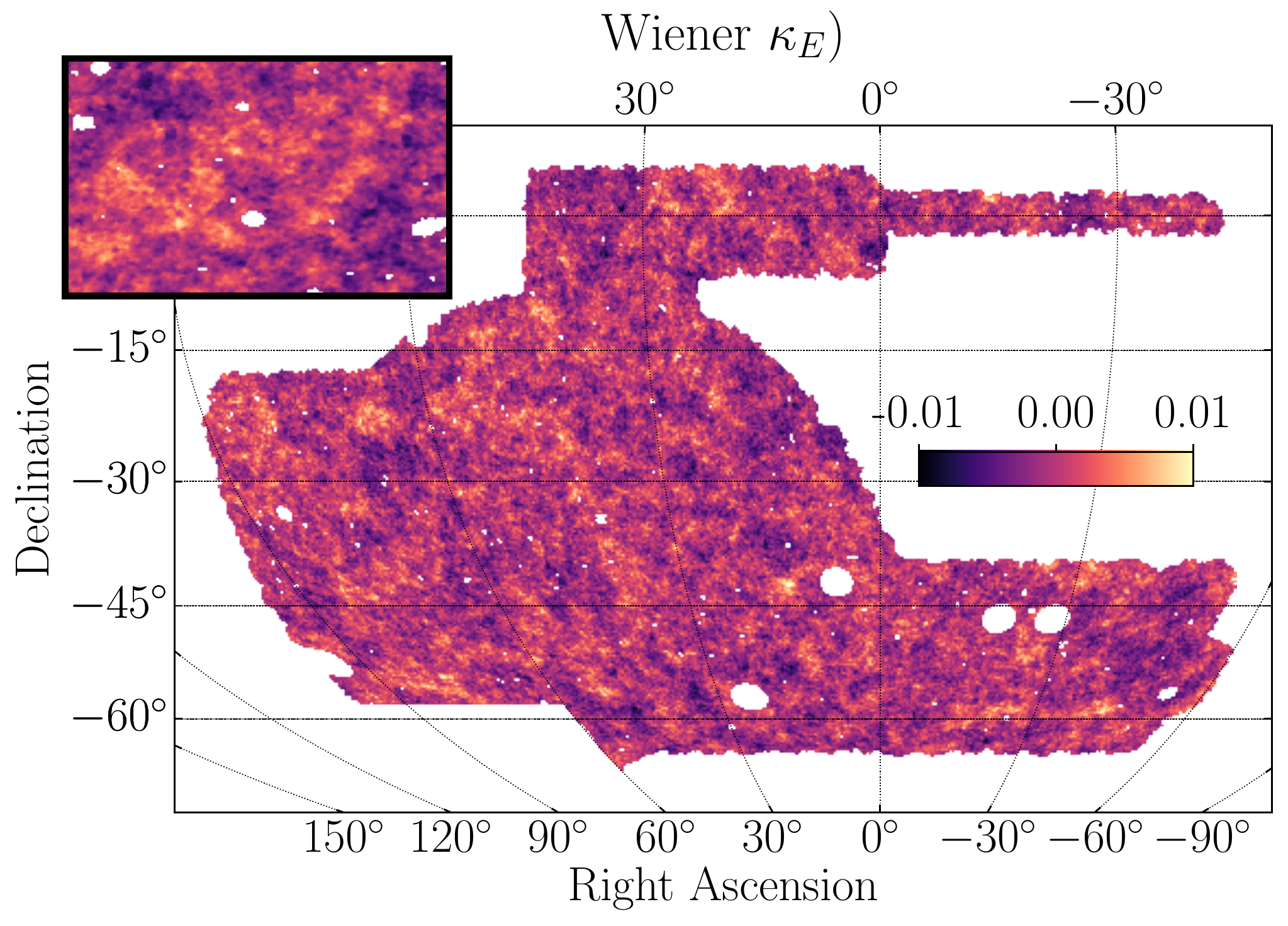}
    \includegraphics[width=0.49\textwidth]{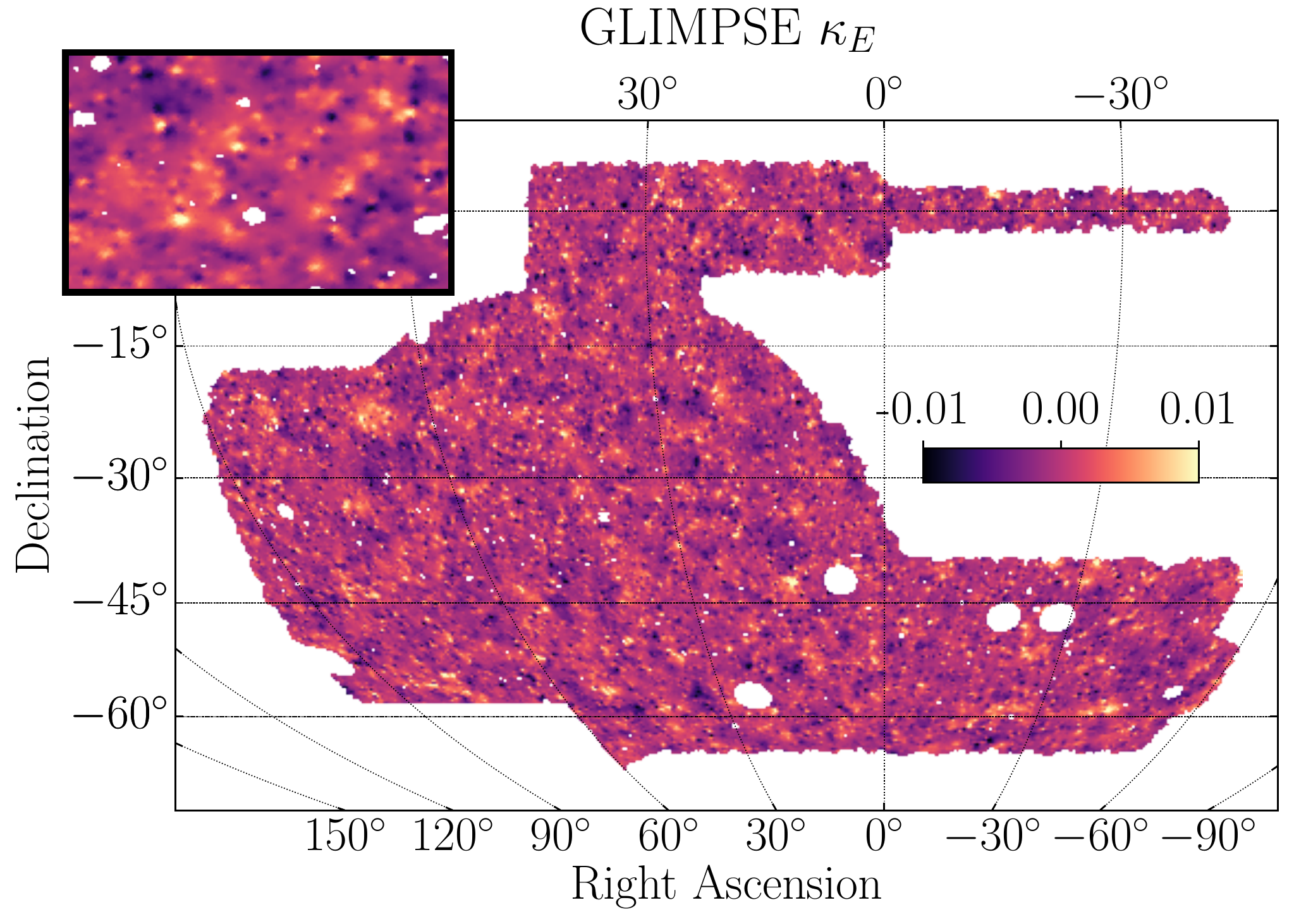} 
    \caption{Simulated DES Y3 weak lensing mass maps, obtained from a noisy realization of the shear field, with different map making methods. \textit{Top left panel}: noisy KS E-mode reconstructed map. \textit{Top right panel}: null B-mode prior method reconstructed map. \textit{Bottom left panel}: Wiener filter reconstructed map. \textit{Bottom right panel}: \glimpse{} reconstructed map. The maps in the top panels have been smoothed at $10~\mathrm{arcmin}$; no further smoothing is applied to the maps showed in the lower panels. \textit{Inset}:  {\small{RA}}$_\mathrm{centre}$, Dec$_\mathrm{centre} = 70^{\circ}, -40^{\circ}$; $\Delta$RA, $\Delta$Dec = 15$^{\circ}$, 10$^{\circ}$.}
    \label{fig:MM_y3_2}
\end{figure*}
In this section we discuss and compare the different mass map methods outlined in Sec.~\ref{sec:massmapping}. To this aim, we use simulated convergence maps and a number of different statistics to test the quality of the reconstruction with respect to the input convergence map available in simulations.  {Here, we only show tests on the maps created using the full shear catalogue.

We do not expect any conclusion drawn in this section to change when considering tomographic maps rather than the full map.} All the maps considered have been converted to \healpix{} \citep{GORSKI2005} maps with $\nside{} = 1024$ (corresponding to a pixel resolution of $3.44~\mathrm{arcmin}$). 

 {As mentioned in the introduction, there is no single comprehensive test for comparative performance between methods. Rather, a number of different tests can be performed, aimed at highlighting the advantages and disadvantages of each method. In particular, Sec. \ref{sec:mask_effects} discusses how different methods deal with mask effects, Sec. \ref{sec:noise_effects} shows the convergence field estimates {in the presence of realistic shape noise} from the different methods when realistic, noisy shear fields are provided as input, while {Secs. \ref{sect:Pearson}--\ref{sect:powerspectrum} show quantitative tests on a number of summary statistics.}} 
{In these tests, whenever meaningful, we varied the parameters of the method (i.e. the $\theta$  parameter for KS and null B-mode prior methods and the $\lambda $ parameter for \glimpse{}). We note that these tests are by no means exhaustive, as other summary statistics could be examined (e.g. higher order statistics, phases, peaks). While we think the tests presented in this section allow us to characterize the advantages and disadvantages of each method, further tests could be performed depending on the particular science application.}

\subsection{ Mask effects}
\label{sec:mask_effects}
{To demonstrate the effects of the mask and missing data, we generate a mock catalogue with no shape noise. Fig.~\ref{fig:MM_y3_1} shows the input true convergence map (\textit{top left}), the KS E-mode reconstruction (\textit{top left}), the KS residual map (\textit{bottom left}), and the KS B-mode map (\textit{bottom right}). The residual is defined as the difference between the input true map and the reconstructed E-mode map. In these figures the maps have been smoothed with a Gaussian kernel with  $\sigma=10~\mathrm{arcmin}$ for visualization.

In the noise-free case all methods other than KS (including Wiener and \glimpse{}) have a null B-mode prior and are thus equivalent. In this noise-free limit, the noise covariance becomes a binary matrix (for the mask) and the signal factors can divide out (although our code implementations of Wiener and \glimpse{} would not be able to deal with this zero limit in practice). The noise-free result is therefore the same for the null B-mode prior method, the Wiener filter, and \glimpse{}.}

From the KS residual map (\textit{bottom left}), where the residual is between the KS idealised case with no shape noise and the truth, we recover most of the features of the input convergence map, except for the part of the map close to the edges of the DES footprint. As discussed in Sec.~\ref{sec:massmapping}, the KS reconstruction is susceptible to mask effects in the case of partial sky coverage, resulting in a non-zero residual map and spurious B-modes (i.e. E-mode leakage).  {The amplitude of the residual map is strongly reduced when a null B-mode prior is applied, as shown in Fig. \ref{fig:residuals_zoomed}. We can also quantify the effect of the null B-mode prior by measuring the power spectra of the recovered maps. In Fig.~\ref{fig:Cl_2} we compare the power spectra of the KS and null B-mode prior maps with the input convergence map power spectrum. The maps have not been smoothed in this comparison. We use the \healpix{} routine \texttt{anafast} to estimate the power spectra of our maps. The power spectra are binned in 20 bins between $\ell=0$ and $\ell = 2048$. Fig.~\ref{fig:Cl_2} clearly shows that the KS method underpredicts the power spectrum at large scales, due to mask effects and E-mode leakage. The null B-mode prior, on the contrary, better recovers the power spectrum at all scales. This holds in the case the spurious B-modes are caused only by mask and edge effects. As all the methods other than KS include a null B-mode prior, these methods are less susceptible to mask effects. }

\subsection{Reconstruction from realistic mock data}
\label{sec:noise_effects}
Fig.~\ref{fig:MM_y3_2} shows the reconstructed maps from the simulated realistic noisy shear catalogue using the four methods for comparison. Again, the KS and the null-B-mode reconstruction have been smoothed at $10~\mathrm{arcmin}$. The \glimpse{} reconstruction uses a sparsity parameter of $\lambda=3$ (discussed below).  {Recall that all the map making methods take into account the noise covariance matrix of the data, thereby characterising the noise amplitude and distribution across the observed area. As a result, all methods naturally take into account inhomogeneities in the noise properties across the DES Y3 footprint.} 

The KS E-mode map is now noticeably degraded compared to the noise-free example (Fig.~\ref{fig:MM_y3_1}). Though the most significant features of the input convergence field can still be spotted by eye, a number of noise-induced small-scale peaks dominate the reconstructed map. The null B-mode prior method map looks similar to the KS E mode map, whereas the impact of noise is reduced in the case of the other methods, due to their signal priors in the map inference process. In particular, the sparsity prior adopted by the \glimpse{} method suppresses the noise enhancing peaky features, which are assumed to be the result of a superposition of spherically symmetric dark matter haloes (a feature that can be noted in the zoomed-in portion of the \glimpse{} map). The noise is also suppressed in the case of the Wiener filter reconstruction, although the map shows fewer peak features compared to the \glimpse{} map. The Wiener method has a prior distribution for which the convergence field is a realization of Gaussian random field, and therefore it is better suited to recover the large-scale structures in the map that have been less affected by non-linear structure collapse. 

\subsection{Pearson correlation coefficient}
\label{sect:Pearson}

\begin{figure*}
\begin{center}
\includegraphics[width=0.9 \textwidth]{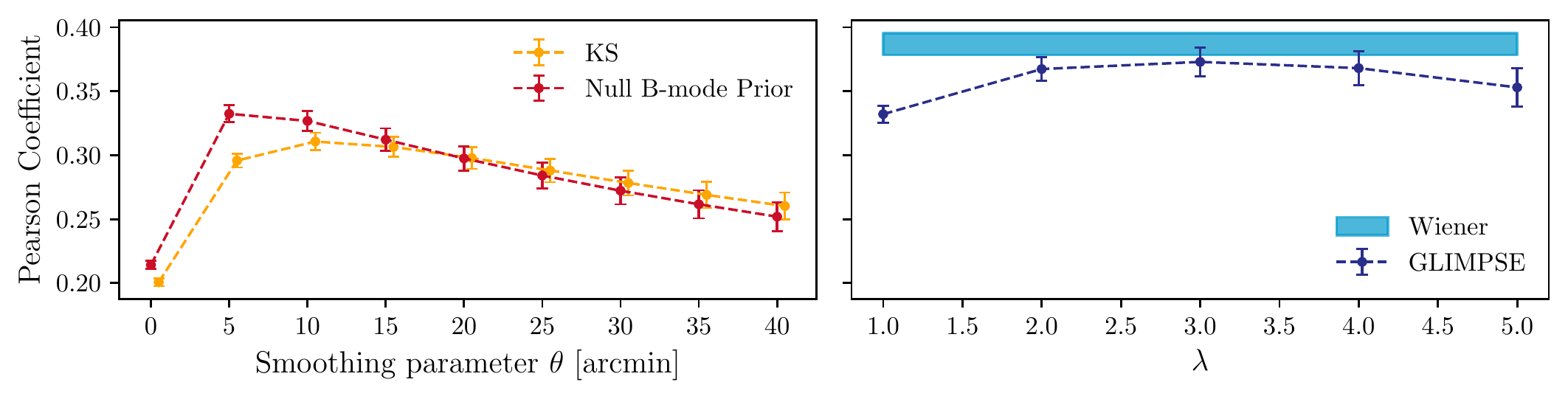} \\
\hspace{-0.1in}\includegraphics[width=0.91 \textwidth]{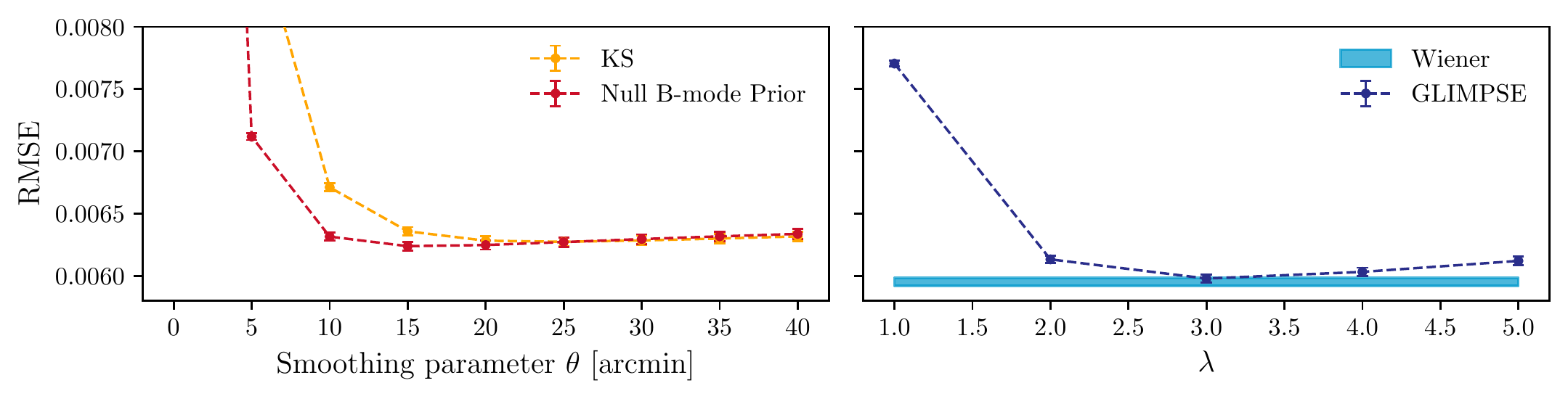}
\end{center}
\caption{Top: Pearson correlation coefficient between the reconstructed map and the true, noiseless convergence map, for the four different mass map methods. When possible, we varied the tuning parameters of the methods. Errors are estimated from jackknife resampling. Bottom: RMSE (see Sec.~\ref{sect:residual} for a definition) for the four different mass map methods. When possible, we varied the tuning parameters of the methods. Errors are estimated from jackknife resampling.}
\label{fig:pearson}
\end{figure*}

{The first statistic we examine is the Pearson correlation coefficient, which quantifies the correlation between the true convergence from simulation and the reconstructed convergence from the simulated mock data catalogue. The Pearson coefficient also reveals the ability of one method to preserve the phases of the convergence field, as it would assume low values if phases were not preserved. The Pearson correlation coefficient, defined for two convergence fields $\kappa_1$ and $\kappa_2$, is given by

\begin{equation}
r_\kappa = \frac{\langle \kappa_1 \kappa_2 \rangle}{ \sqrt{\langle\kappa^2_1 \rangle} \sqrt{\langle \kappa^2_2 \rangle}} \ \ ,
\end{equation} 

\noindent where $\langle \kappa_1 \kappa_2 \rangle$ is the sample covariance estimated using pixel values of $\kappa_1$ and  $\kappa_2$.

In this case, we compute the Pearson correlation coefficient between the true simulated convergence map and the reconstructed E-mode convergence map. The results are shown in Fig.~\ref{fig:pearson}. In general, the closer to unity the Pearson coefficient value, the better the reconstruction.}

For KS and the null B-mode prior methods the smoothing parameter of the Gaussian kernel $\sigma$ was varied, while for \glimpse{} we varied the sparsity parameter $\lambda$. Recall that in our implementation of the null B-mode prior method the map is recursively smoothed at every iteration of the algorithm, so that the final map is slightly smoother than if it were smoothed only at the end of the iterative procedure. This means that in practice a given value of the smoothing parameter $\theta$ for the null B-mode prior method should be compared to a slightly larger value $\theta$ for the KS method.

The effect of the tuning parameter for the null B-mode prior method is similar to KS, although the former method performs slightly better at small smoothing parameter values. The KS and null B-mode prior methods maximise the Pearson coefficient at $10\ \mathrm{and}\ 5~\mathrm{arcmin}$ of smoothing respectively. This is due the small angular scales being shape noise dominated, with $5-10~\mathrm{arcmin}$ corresponding to the scale where the amplitude of shape noise is comparable to the amplitude of the signal. One can interpret this as the smoothing up to $5-10~\mathrm{arcmin}$ removing more small-scale noise-induced spurious structures than true signal. A different shape noise contribution (or, equivalently, a different data set) would change this scale; in the limit of no shape noise, the optimal scale would be the smallest scale allowed by the pixelization scheme. The null B-mode prior method performs slightly better than KS at small $\theta$ because of the extra regularization (i.e. smoothing) performed at every step of the iterative algorithm; this further suppresses noise, improving the Pearson coefficient at small scales.

For \glimpse{}, the level of suppression of the shape noise is controlled by the sparsity coefficient $\lambda$, for which we found $\lambda = 3$ to optimize the Pearson correlation coefficient. The Wiener filter has no free parameters in our implementation provided the fiducial power spectrum is assumed. Both \glimpse{} and the Wiener filter outperform standard KS and null B-mode prior methods, delivering a higher Pearson coefficient.

\subsection{RMSE}\label{sect:residual}

The second statistic we examine is the root-mean-square error (RMSE) of the residuals, defined to be
\begin{equation}
{\rm RMSE}(\kappa^{\rm truth},\kappa^{\rm recon}) \equiv \sqrt{\frac{1}{N}\sum_{p=1}^N \Delta \kappa_p^2}, 
\end{equation}
where $N$ is the number of pixels and $\Delta \kappa_p$ is the difference between the reconstructed map and the true map in pixel $p$. Again, we only consider E-mode maps and maps recovered from noisy estimates of the shear field. The results are shown in Fig.~\ref{fig:pearson}. In general, the closer to zero the RMSE, the better the reconstruction. The RMSE reveals the ability of one method to preserve the phases and the amplitude of the convergence field.

The results from this test match those from the Pearson coefficient test. The null B-mode prior method shows a similar trend to the KS method, although it is characterized  by a smaller RMSE at small scales. The \glimpse{} and Wiener methods perform better (i.e. the RMSE is closer to zero) than standard KS and the null B-mode prior methods.

For KS and the null B-mode prior methods the RMSE is reduced strongly with smoothing, indicating that the variance at small scales is completely dominated by shape noise, reaching a minimum after smoothing the reconstructed maps at $10-20~\mathrm{arcmin}$. We note that the minimum of the RMSE signal and the maximum of the Pearson coefficient for these two maps are at a similar smoothing parameter value (even though the value does not need to be exactly the same). For these two methods, the RMSE should converge at very large smoothing parameter values (larger than those showed here) to the RMSE of the original field, as the reconstructed map signal goes to zero. Similar to the Pearson coefficient case, the null B-mode prior method has a smaller RMSE compared to KS at small scales, due to the extra noise suppression of the algorithm.

The \glimpse{} and Wiener methods have a significantly smaller RMSE compared to KS, meaning the reconstructed \glimpse{} and Wiener maps map is more accurate than KS on the pixel level. For \glimpse{} the minimum RMSE is reached for a sparsity parameter $\lambda = 3$, the same value that maximises the Pearson coefficient.

\subsection{Power spectra}\label{sect:powerspectrum}

\begin{figure*}
\begin{center}
\includegraphics[width=0.99 \textwidth]{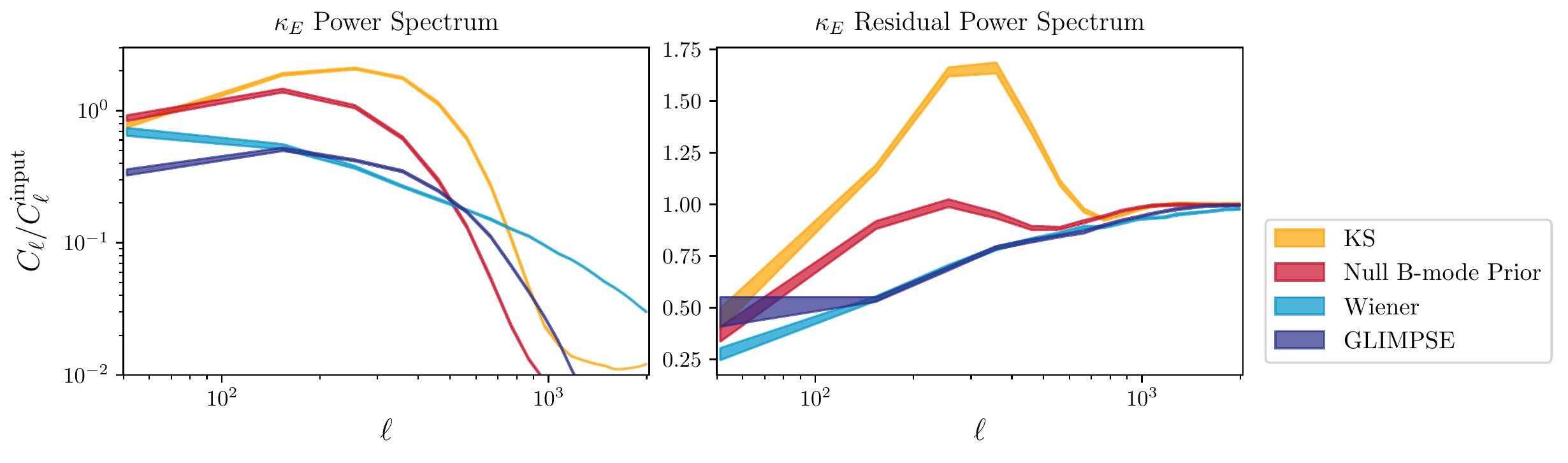}
\end{center}
\vspace{-0.1in}
\caption{ {Simulations.} \textit{Left panel}: power spectrum of the reconstructed maps obtained from a noisy realization of the shear field with respect to the power spectrum of the input convergence field. For the KS  and  the  null B-mode  prior methods,  we  considered the  maps  with  $10~\mathrm{arcmin}$  smoothing; for the \glimpse{} method, we considered the map obtained with sparsity parameter $\lambda=3$. As discussed in Sec.~\ref{sec:prop}, the power of the \textit{maximum a posteriori} estimates will not match the power of the truth, and is expected to be reduced. For the Wiener filter, this reduction is known analytically as a function of signal and noise covariance. \textit{Right panel}: power spectrum of the residual map, defined as the difference between the recovered map and the input convergence field.}
\label{fig:Cl}
\end{figure*}

We now examine, for each method, the power spectrum of the residual map (defined to be the difference between the reconstructed map and the input convergence map) and the power spectrum of the reconstructed map. Recall that the reconstruction $\hat{\boldsymbol{\kappa}}$ is a \textit{maximum a posteriori} estimate, so the power spectrum of $\hat{\boldsymbol{\kappa}}$ is not expected to match the power spectrum of the underlying field (Sec.~\ref{sec:prop}).

The differences between power spectra highlight the effect of different priors on the \textit{maximum a posteriori} reconstruction, whereas the residual map power spectra show at which scales the recovered maps are most similar to the input convergence field. For these tests, we use the maps recovered from a noisy version of the shear field. We use the \healpix{} routine \texttt{anafast} to estimate the power spectra of our maps. The power spectra are binned in 20 bins between $\ell=0$ and $\ell = 2048$. For the KS and the null B-modes prior methods, we considered the maps with $10~\mathrm{arcmin}$ smoothing; for the \glimpse{} method, we considered the map obtained with sparsity parameter $\lambda=3$.

 The left panel of Fig.~\ref{fig:Cl} shows the power spectra of the maps compared to the power spectrum of the input convergence field. There is a clear signal suppression at small scales and high multipoles; this is a consequence of the priors implemented by the different methods to reduce the impact of noise (which dominates the small-scale regime). The KS and the null B-modes prior methods show similar behaviour, as they implement similar priors; however, the null B-mode prior method suppresses the small-scale signal slightly more compared to KS. In general, none of the methods reproduce the correct amplitude of the input theory power spectra; this is to be expected with point-estimate reconstructions of the map (Sec.~\ref{sec:prop}).

 The right panel of Fig.~\ref{fig:Cl} shows the power spectra of the residual maps. At large scales the Wiener map shows the smallest amplitude, indicating that it performs best at reproducing the large-scale pattern of the convergence field. For Wiener and \glimpse{} maps, the residuals steadily increase at larger multipoles; indeed, none of the methods is able to recover the small-scale information. Besides this main trend, the KS and null B-mode prior maps also show an increment in the residual map power spectrum around $\ell \sim 300$. The smoothing prior is not able to reduce the impact of shape noise at these scales, causing the residual map power spectrum to increase substantially. This shows that the Wiener and \glimpse{} methods are indeed better than the KS and null B-mode prior methods at recovering intermediate scales.

\subsection{Convergence one-point distribution and recovery of the input convergence pixel values}\label{sect:pdf}
\begin{figure*}
\begin{center}
\includegraphics[width=0.98 \textwidth]{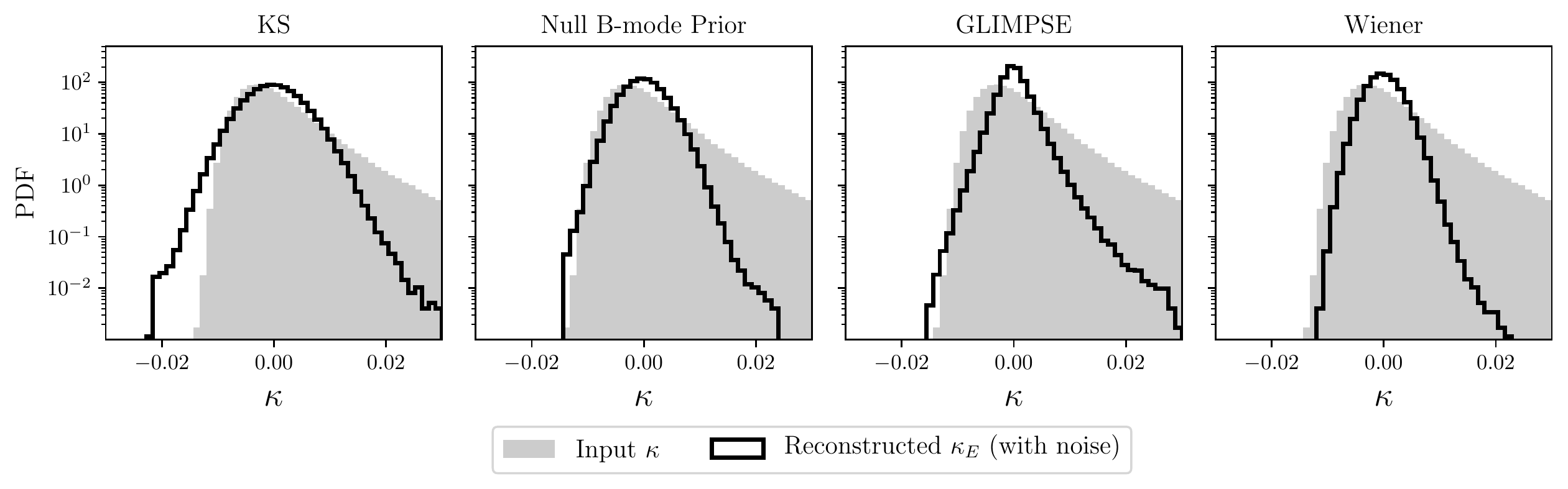}
\end{center}
\vspace{-0.1in}
\caption{PDFs (one-point distributions) for the different
map reconstruction methods, obtained from a simulated, noisy realization of the shear field. The grey shaded histogram in each panel is the PDF of the true, input convergence field.}
\label{fig:PDF}
\end{figure*}

\begin{figure*}
\begin{center}
\includegraphics[width=0.98 \textwidth]{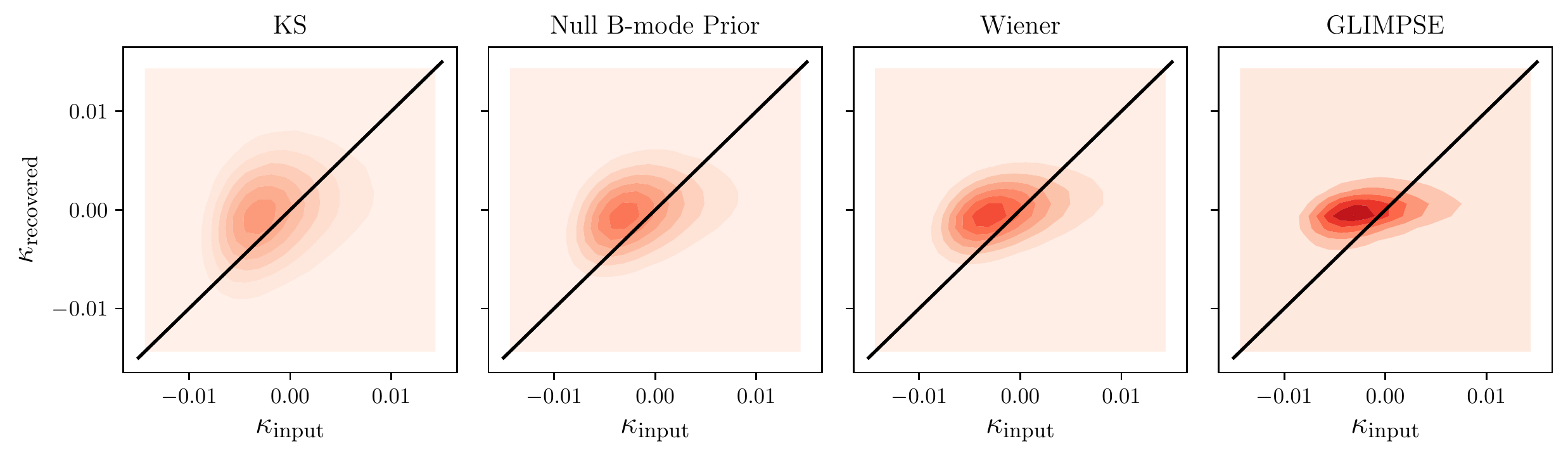}
\end{center}
\vspace{-0.1in}
\caption{Density plots showing the relation between the values of the pixels of the recovered maps and the input convergence field map. A map that perfectly recovered the truth would have a density plot that followed the black solid line. All of the density plots are normalised. The linear correlation between variables shown in this plot is quantified by the Pearson correlation coefficient discussed in section \ref{sect:Pearson}. As discussed in Sec.~\ref{sec:prop}, the pixel variance of the as \textit{maximum a posteriori} estimates will not be equal to pixel variance of the truth, and is expected to be reduced.} 
\label{fig:1point}
\end{figure*}

In Fig.~\ref{fig:PDF} we show the one-point distribution function (PDF) of the convergence field. For KS and the null B-mode prior reconstructions we considered maps with $10~\mathrm{arcmin}$ smoothing, and we used $\lambda=3$ for \glimpse{}.

The PDFs of the pixel values of the reconstructed maps are not identical to those of the input. This is expected. As all reconstructions are a \textit{maximum a posteriori} estimate of the underlying convergence field, the variance (and possibly higher-order moments) of the reconstructed map will be suppressed. The asymmetric distributions are a sign that the recovered map is not dominated by noise, whose PDF is completely symmetric.

We also show in Fig.~\ref{fig:1point} density plots illustrating the relation between the values of the pixels of the recovered maps and those of the input convergence map. For a perfect reconstruction, the density plots would look like a straight, diagonal line (the black line in the Figure). In general, it can be noted that the values of the pixels of the recovered maps scatter more around zero than the values of the pixels of the input map. This is a consequence of the noise; however, as already noted in Fig.~\ref{fig:PDF}, the density plots not being perfectly symmetric means that the maps are not dominated by noise. Generally, pixels with negative (positive) values in the recovered maps are also associated to the ones with negative (positive) values in the input convergence map, although with a large scatter. The scatter is larger for pixels with positive values, due to the long positive tail of the convergence PDF.

The density plots for the Wiener filter and \glimpse{} maps are tighter, whereas KS and null B-mode prior method show a larger scatter. The density plots convey the same information encoded by the RMSE: a higher (lower) RMSE value is associated to a tighter (broader) density plot around the black diagonal line in Fig.~\ref{fig:1point}.

\section{Application to data}
\label{sec:data_maps}

\subsection{Map reconstruction}

{  In this section we present the reconstructed mass maps using DES Y3 weak lensing data. We show only maps created using the full catalogue. We also created maps for the four tomographic bins; they are not shown here, but they will be made publicly available following publication at} \url{https://des.ncsa.illinois.edu/releases}.

\begin{figure*}
\begin{center}
    \includegraphics[width=0.49\textwidth]{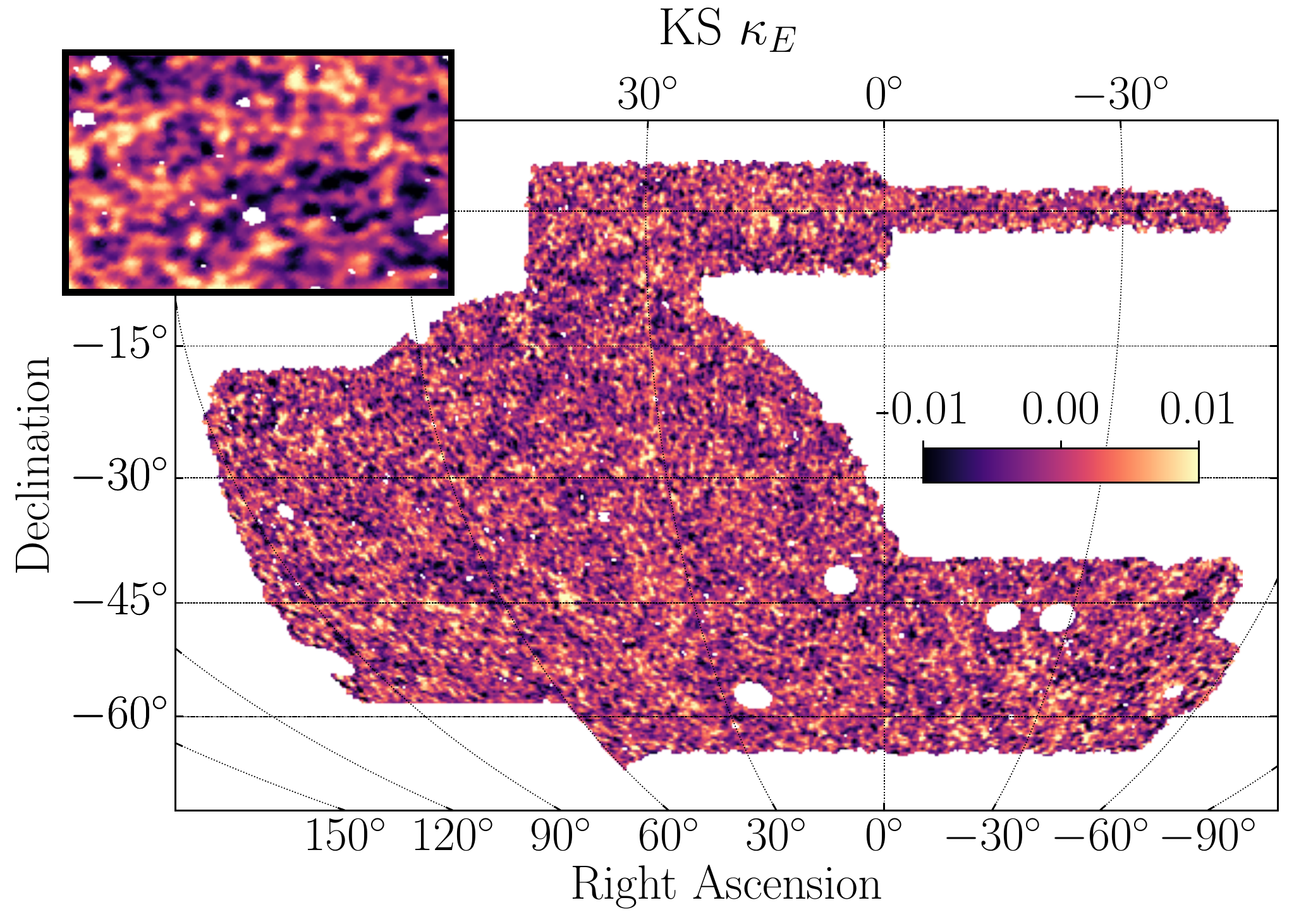}
    \includegraphics[width=0.49\textwidth]{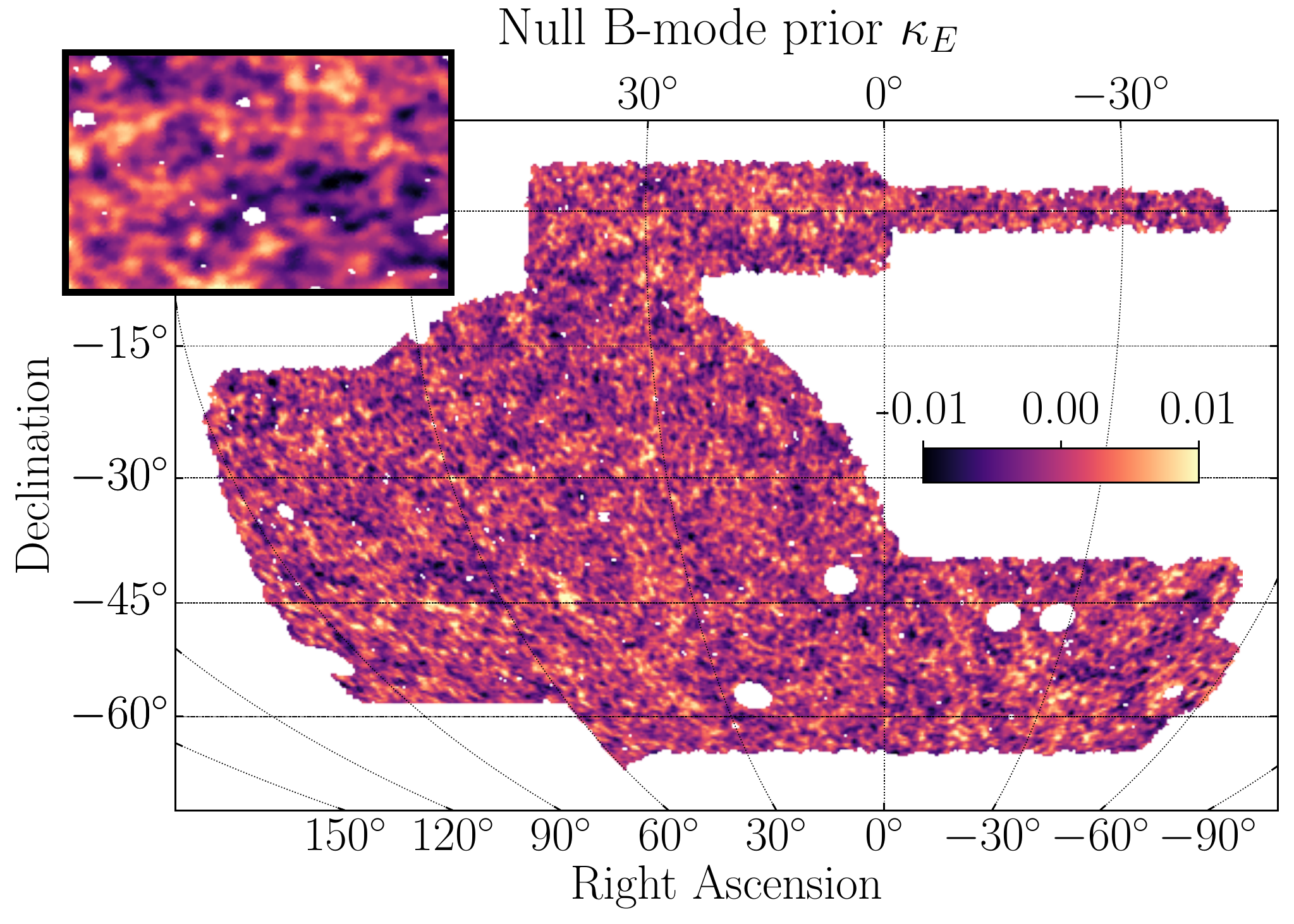} \\
    \includegraphics[width=0.49\textwidth]{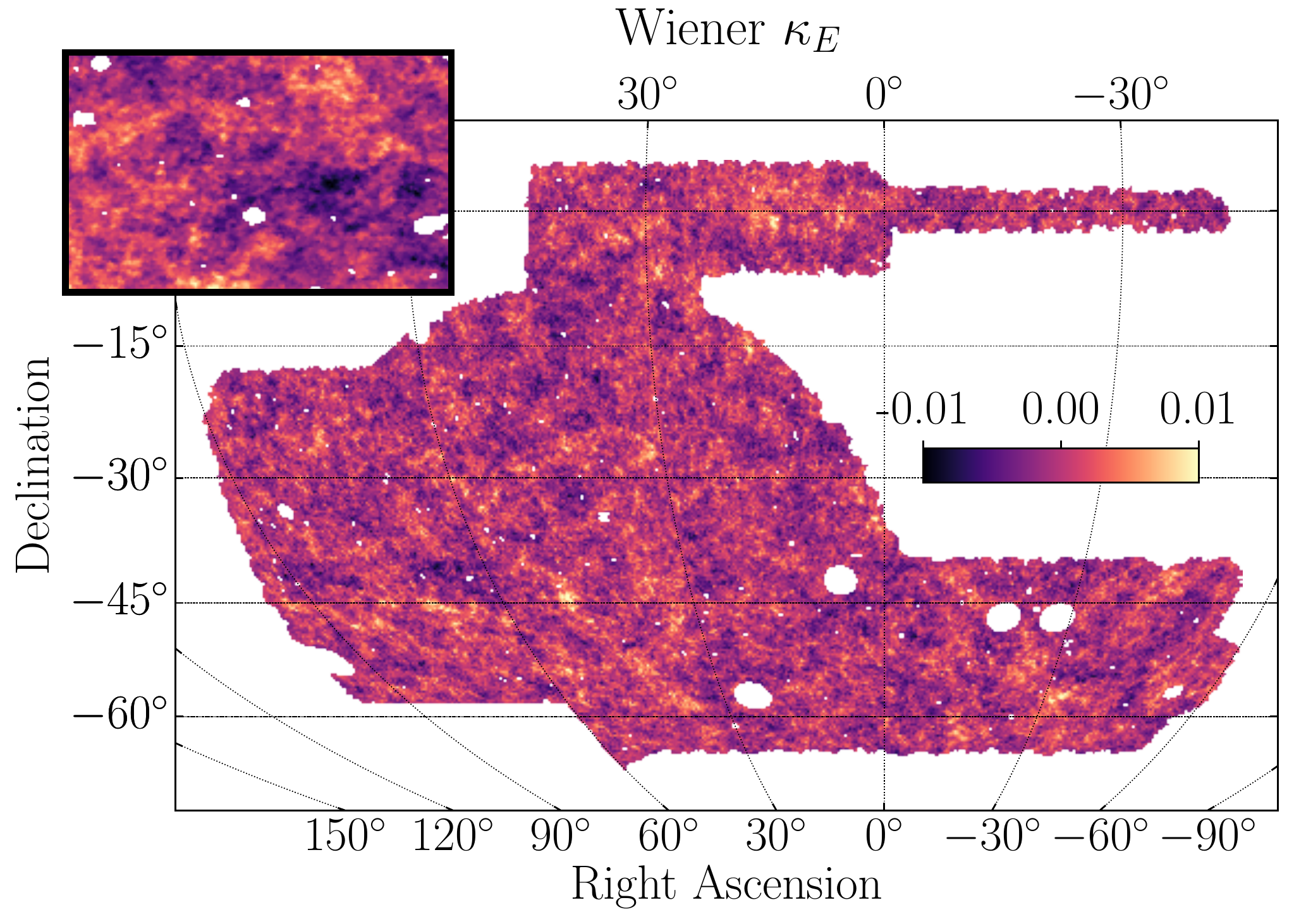}
    \includegraphics[width=0.49\textwidth]{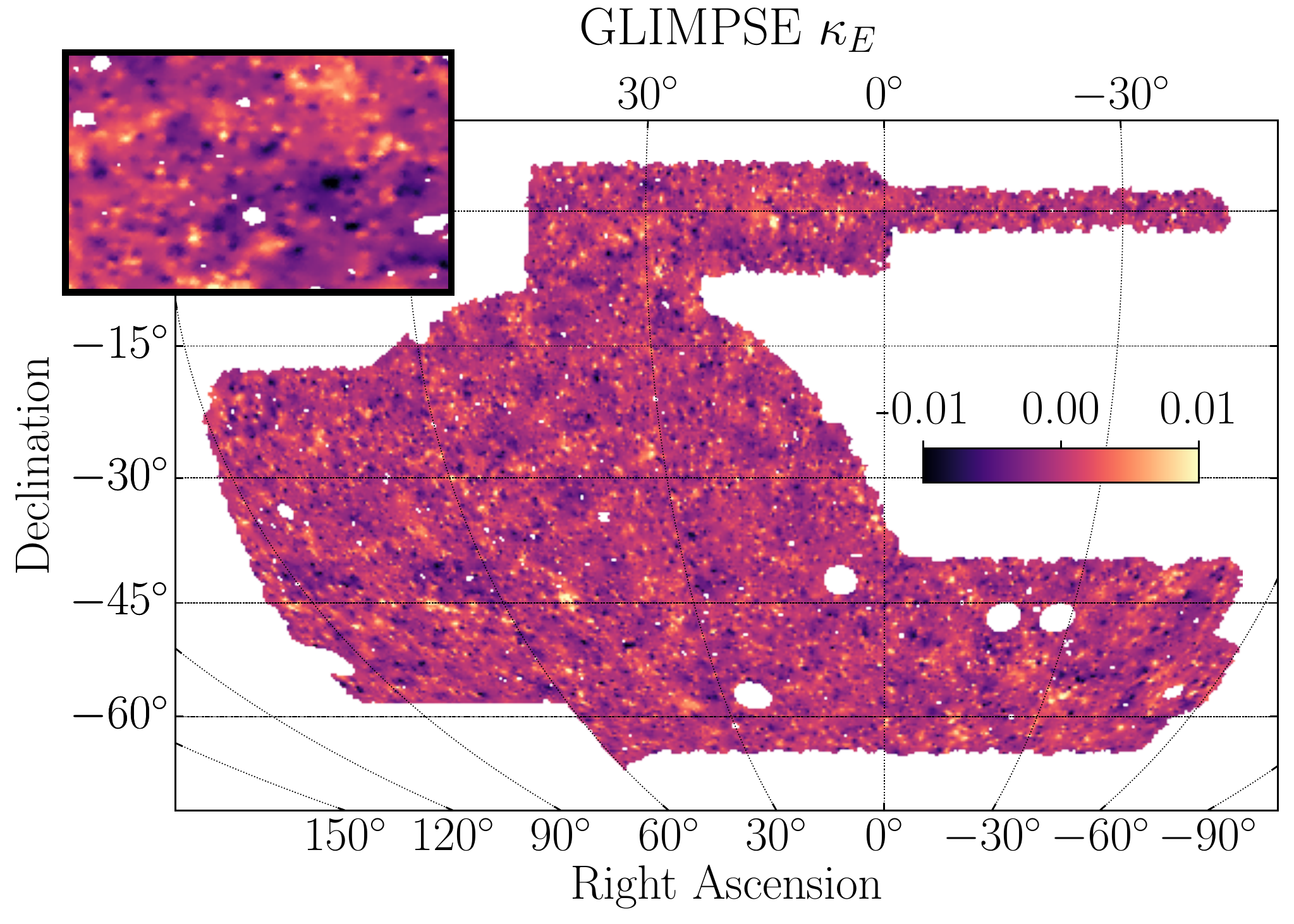} 
\end{center}
\caption{\texttt{METACALIBRATION} DES Y3 weak lensing mass maps, obtained from the official DES Y3 shear catalogue and created using different map making methods. \textit{Top left panel}: noisy KS E-mode map \textit{Top right panel}: E-mode map obtained with the null B-mode prior method. \textit{Bottom left panel}: E-mode Wiener filter map. \textit{Bottom right panel}: E-mode \glimpse{} map. The maps in the top panels have been smoothed at $10~\mathrm{arcmin}$; no further smoothing is applied to the maps showed in the lower panels. \textit{Inset}:  {\small{RA}}$_\mathrm{centre}$, Dec$_\mathrm{centre} = 70^{\circ}, -40^{\circ}$; $\Delta$RA, $\Delta$Dec = 15$^{\circ}$, 10$^{\circ}$.}
\label{fig:fiducial_map}
\end{figure*}

Fig.~\ref{fig:fiducial_map} shows the four maps obtained with the KS, null B-mode prior, Wiener filter and \glimpse{} methods, obtained from the \texttt{METACALIBRATION} catalogue. We recall that these maps have been obtained applying the \texttt{METACALIBRATION} response correction and the inverse variance weights, as explained in Sec.~\ref{sec:data_sims}. The maps obtained with the different methods visually show the same differences as the ones obtained in simulations (Fig.~\ref{fig:MM_y3_2}), with the Wiener and \glimpse{} maps particularly suppressing the noise thanks to their priors.

\begin{figure*}
\begin{center}
    \includegraphics[width=0.47\textwidth]{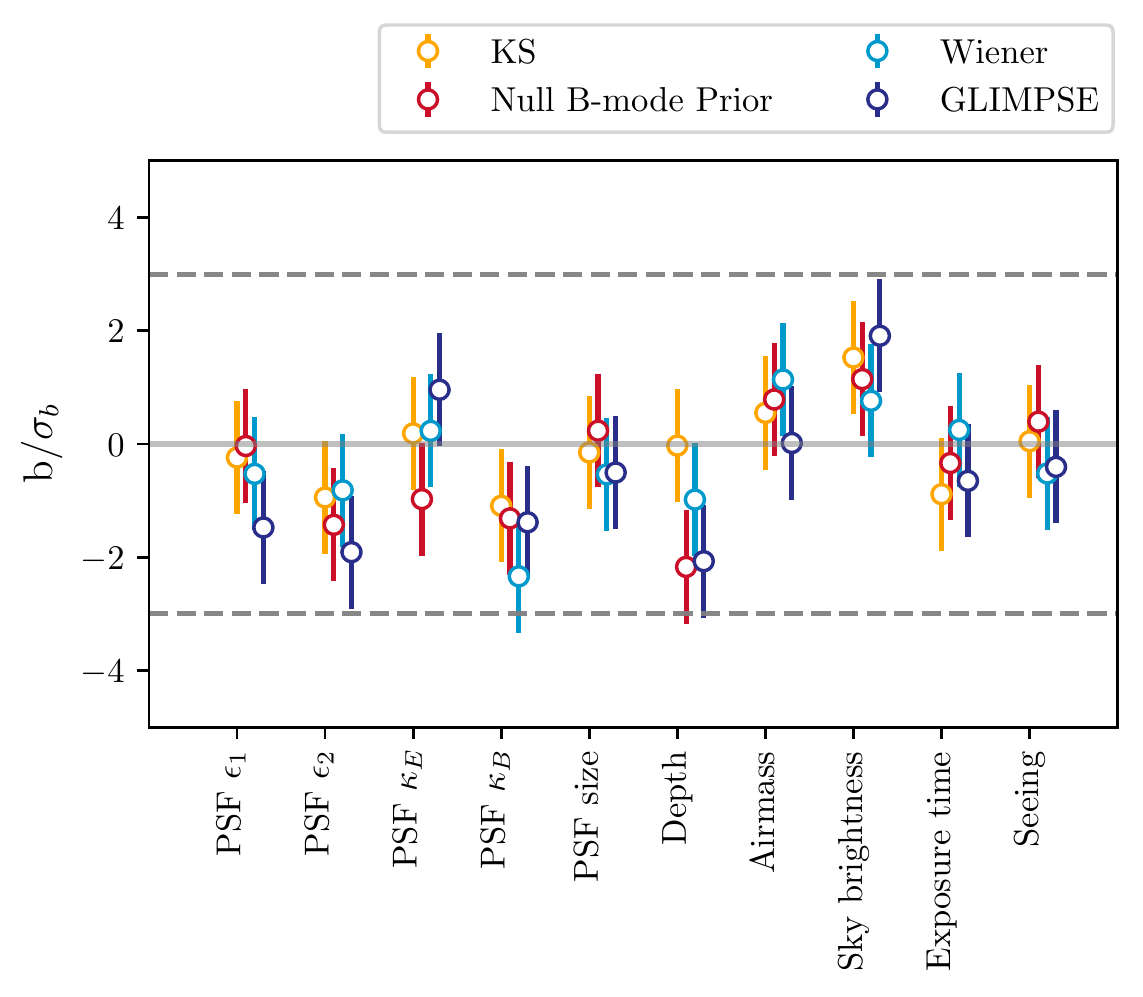}
    \includegraphics[width=0.49\textwidth]{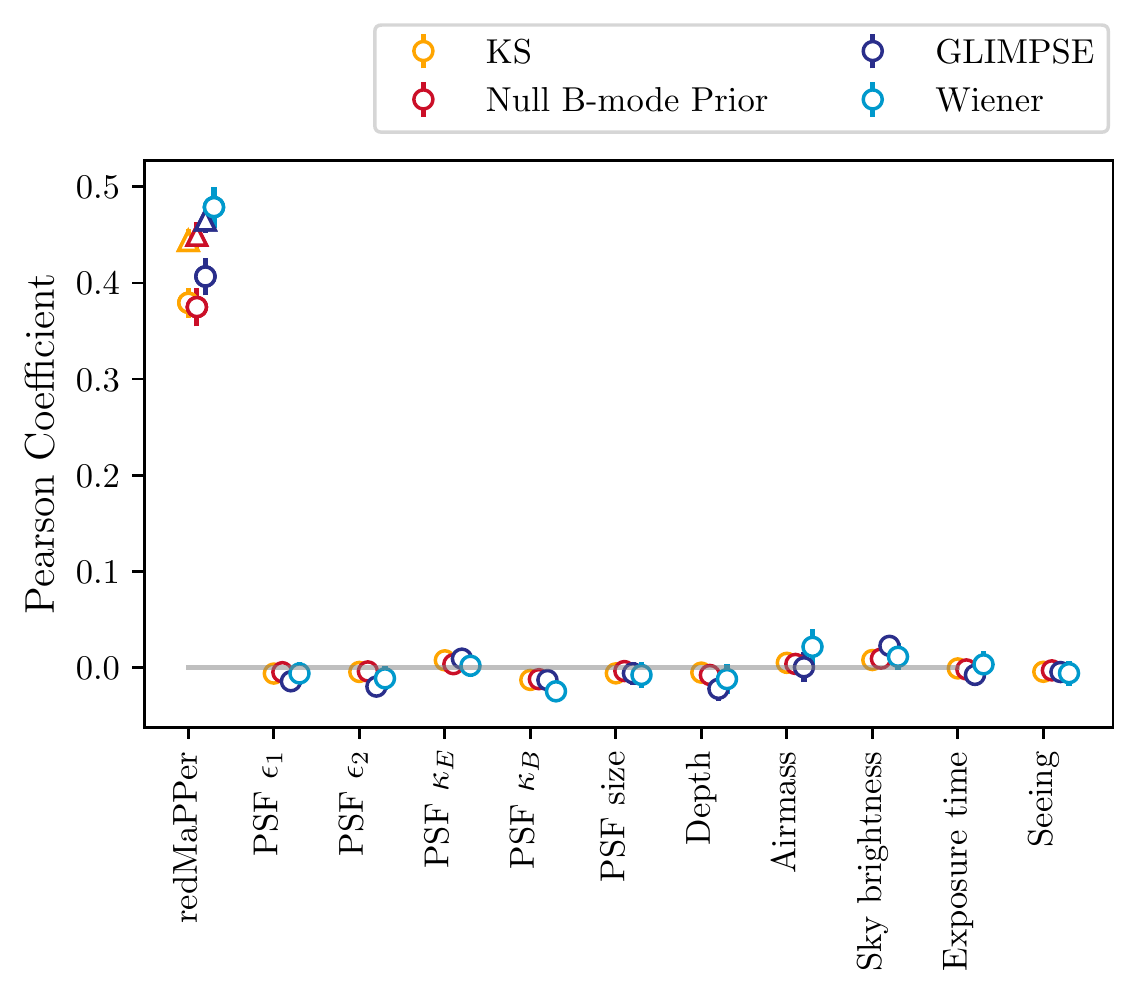}

\end{center}
\caption{\textit{Left panel}: Best fit values for the coefficient of the relation $\kappa_{\rm E} = b M^{\rm S}$ for a given systematic map $S$. The values of the slopes are shown for different tomographic bins, and the uncertainties are estimated through jackknife resampling.  \textit{Right panel}: Pearson coefficient between the recovered convergence map and the systematic maps $S$. Uncertainties are estimated through jackknife resampling. When applicable, systematic maps are considered in the i-band. For the redMaPPer cluster correlation (\textit{right panel}) we also show the result for different tuning parameters (see text for details) that are shown with with triangle markers.}
\label{fig:syst}
\end{figure*}

\subsection{Systematic error tests}

We perform a number of tests on the recovered maps. We first test if any spurious correlation exists between our maps and quantities that are not expected to correlate with the convergence maps. The shear catalogue used to produce the mass maps have been largely tested in \cite*{y3-shapecatalog}, but the potential correlation between convergence maps and systematic errors was not investigated there. We therefore consider a number of catalogue and observational properties as potential systematic errors, in a fashion similar to what was done in \cite*{y3-shapecatalog}. In particular, we consider the two components of the point-spread-function (PSF) ellipticity at the galaxy position (${\rm PSF}_1$,  ${\rm PSF}_2$), their E and B-modes maps (${\rm PSF}_E$, ${\rm PSF}_B$), and the size of the PSF ($T_{\rm PSF}$). As observing condition properties, we consider mean air-mass, mean brightness, mean magnitude limit (depth), mean exposure time, and mean seeing (all in the i-band).

A few maps were considered in the shear catalogue tests and so are excluded here. For example, we do not include the signal-to-noise ratio maps among the systematic maps, as we actually expect to measure a signal (indeed, overdense regions of the sky should be populated by red elliptical galaxies with high signal-to-noise). Similarly, we expect (and measure) at high significance a correlation between galaxy colours and our mass maps.

\begin{figure*}
\begin{center}
    \includegraphics[width=0.8\textwidth]{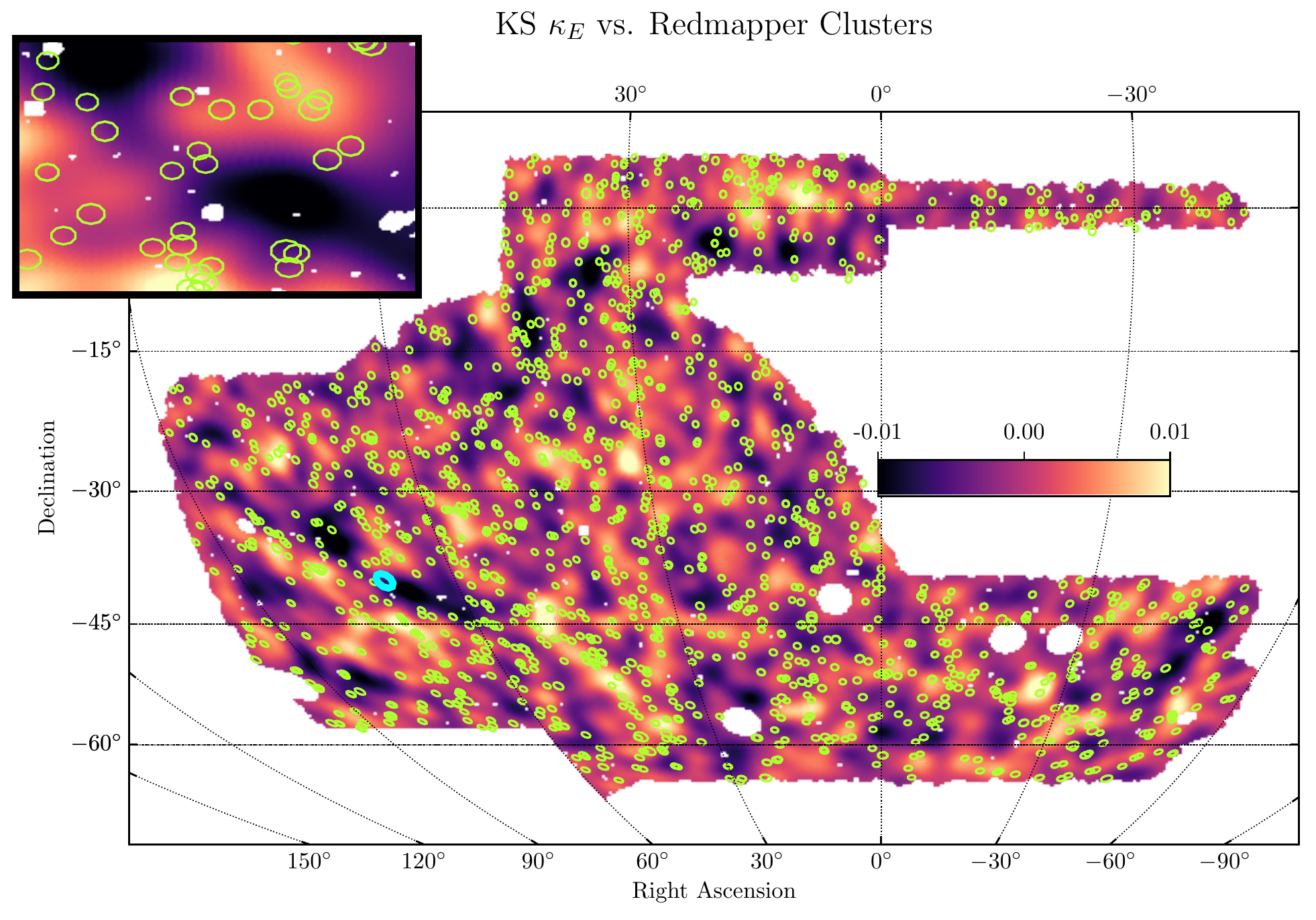} 
\end{center}
\vspace{-0.1in}
\caption{\texttt{METACALIBRATION} DES Y3 weak lensing mass maps using galaxies in the third redshift bin (see Figure~\ref{fig:Nz_sims_data}), obtained with the KS method, with redMaPPer clusters in the redshift range $0.3<z<0.5$ (green circles) superimposed. In the wide field, we randomly selected a subsample of the clusters with richness $\lambda_{\rm RM}>35$; for the small inset, we zoom in on the (randomly chosen) location $(\textrm{RA}, \textrm{Dec})=(70^{\circ}, -40^{\circ})$ (cyan marker on the large map). The circles are centred at the cluster centre, with the size of the circles scaling with the mass (richness) of the clusters. Visually, the clusters coincide with the high $\kappa$ regions and avoid the low $\kappa$ regions. The $\kappa$ map is smoothed by a 1~deg Gaussian filter to highlight large-scale features. }
\label{fig:fiducial_map_cluster}
\end{figure*}

We follow \cite{Chang2018} and create (using mean-subtracted values) a systematic map $M^{\rm S}$ for each of the systematic errors. We first assume a linear dependence between the convergence maps and the systematic maps:
\begin{equation}
\kappa_{\rm E} = b M^{\rm S}.
\end{equation}
We fit all the pixel values of the convergence maps assuming such a linear relationship with the systematic maps. We show the measured coefficient for each of these systematic maps in the left panel of Fig.~\ref{fig:syst}. Errors are estimated using jackknife errors. We do not find any particularly significant correlation; individually, the coefficients are measured with a significance smaller than $3 \sigma$. The overall $\chi^2$ of the null hypothesis (considering the correlations among the 10 systematic maps considered here) is 6, 12, 10, and 17  for 10 $d.o.f.$, for KS, null B-mode prior, Wiener, and \glimpse{} respectively, indicating compatibility with no significant dependence on systematic errors. We also compute the Pearson coefficient between the convergence maps and the systematic maps; results are shown in the right panel of Fig.~\ref{fig:syst} (note that in the same Figure we also show the Pearson coefficient with redMaPPer clusters, discussed in the next section). The main difference with the linear fit is that the Pearson coefficient does not assume \textit{a priori} any relation between the convergence maps and systematic maps. Again, we do not find any strong evidence of systematic contamination, with the $\chi^2$ of the null hypothesis being 5, 5, 7, and 10 for 10 $d.o.f.$, for KS, null B-mode prior, Wiener, and \glimpse{} respectively.

\subsection{Structures in the reconstructed maps}

\subsubsection{Galaxy cluster distribution}

For obvious reasons the true convergence map is not available in data; nevertheless we can check that the reconstructed mass maps probe the foreground matter density field by correlating them with a sample of other tracers. For visualization purposes, we show in Fig.~\ref{fig:fiducial_map_cluster} the \glimpse{} map with a few redMaPPer clusters superimposed.

{  From Fig.~\ref{fig:fiducial_map_cluster} we can see that clusters tend to populate the densest regions in the reconstructed convergence map and avoid the regions with negative convergence signal.}

{  We also report in Fig.~\ref{fig:syst} the Pearson coefficient between the maps and the effective richness of redMaPPer clusters at $z<0.6$. In particular, we follow \cite{jeffrey2018} and define an effective lensed cluster richness $\lambda_R^{\rm eff}$:}
\begin{equation}
\lambda_R^{\rm eff}  = \lambda_R  \frac{p(\chi) \chi }{a(\chi)},
\end{equation}
{where $\lambda_R$ is the redMaPPer cluster richness, $a(\chi)$ is the scale factor evaluated at the comoving distance to a given cluster $\chi$, and $p(\chi)$ is the lensing efficiency, defined as $p(\chi) = \int_{\chi}^{+\infty} d\chi' n(\chi') \frac{\chi'-\chi}{\chi'}$, with $n(\chi')$ the redshift distribution of the source galaxies used to create the mass maps as a function of comoving density. The effective richness is then normalised to the mean of the effective richness of all clusters considered. For all the maps, the measured Pearson coefficient shown in Fig.~\ref{fig:syst} is significantly larger than 0, showing how the recovered maps successfully trace the foreground matter density field. Again, we use parameter value $\theta= 10$ {arcmin} for the KS and the null B-mode prior reconstruction and $\lambda=3$ for \glimpse{} by default. For the redMaPPer result in the \textit{right panel} of~Fig.~\ref{fig:syst}, we also plot $\theta = 5$ arcmin and $\lambda = 1$ (the triangular figure markers), which were shown to improve the correlation for these maps.}


\begin{figure}
\begin{center}
\includegraphics[width=88mm]{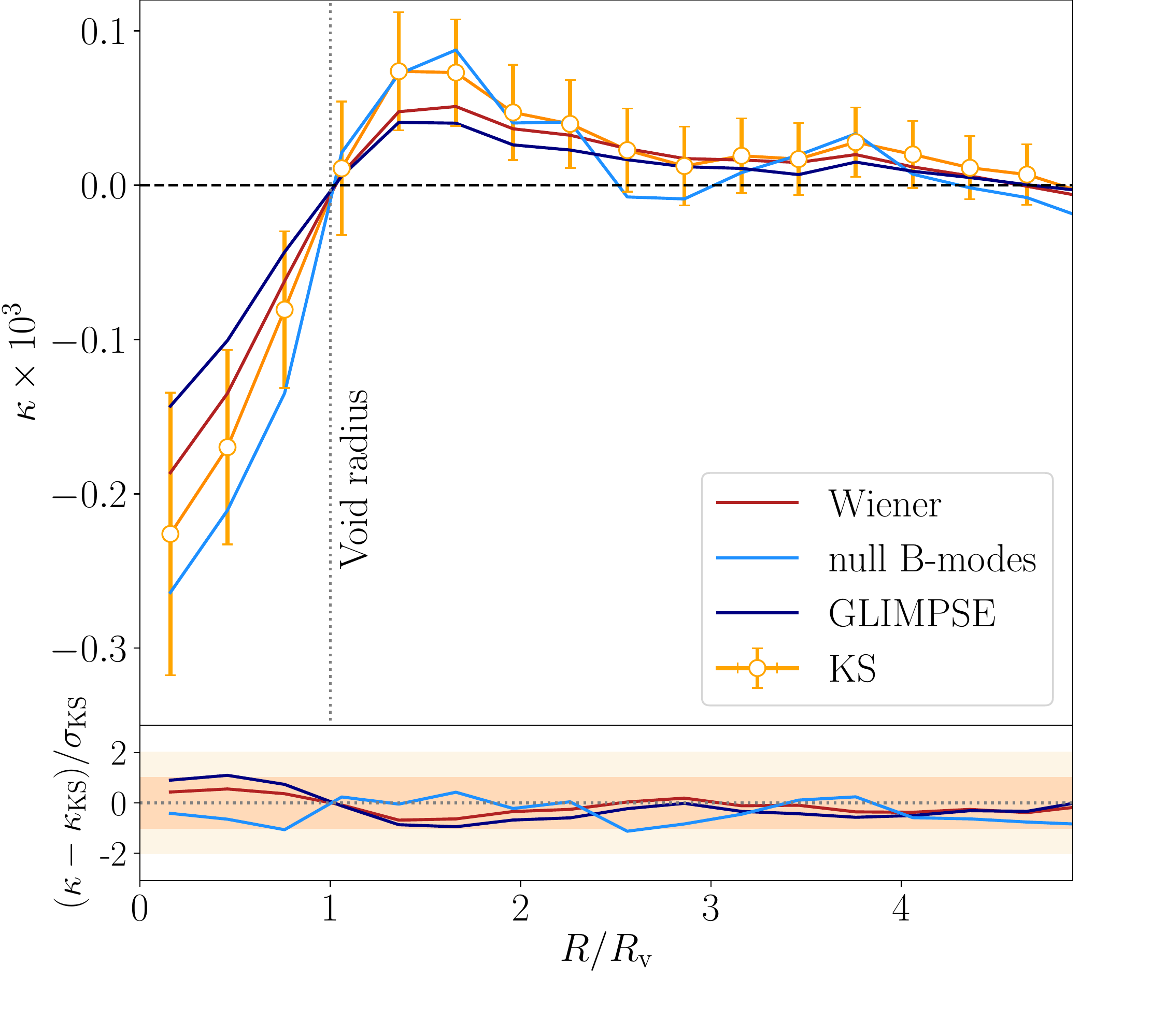}
\end{center}
\caption{\textit{Top panel:} Void imprints on the DES mass maps. \textit{Bottom panel:} Differences of signals measured from different mass maps, relative to the KS results and errors (shaded ranges are $1\sigma$ and $2\sigma$ about the KS signal).}
\label{fig:void_profile} 
\end{figure}

\subsubsection{Cosmic void imprints}

{Cosmic voids are an increasingly favoured cosmic probe and have now already been successfully used to extract cosmological information \citep[for a recent overview see][]{voids_white_paper}. We expect these large lower-density regions in the cosmic web to display a typical imprinting in the convergence signal when cross-correlated with weak lensing mass maps \citep[for previous results from DES Y1 data see][]{Chang2018}.

We create a catalogue of so-called `2D voids' \citep{Sanchez2017} from the DES Y3 \redmagic{} \citep{rozo} photometric redshift data set by searching for projected underdensities in tomographic slices of the galaxy catalogue. On average, these tunnel-like voids correspond to density minima that are compensated by an overdense zone in their surroundings. With this simple approach, we detect $3,222$ voids in the DES Y3 data set, which are larger on average, although also less underdense, than most voids from other void finders \citep[see e.g.][]{Fang2019}. They certainly are useful tools in void lensing studies \citep{Davies2018} and they have been widely used in previous DES analyses \citep[see e.g.][]{vielzeuf,Kovacs2017,Kovacs2019, Fang2019}.

The lensing imprint of typical individual voids is expected to be undetectable \citep{Amendola1999}. Therefore, after selecting our void sample, we follow a stacking method to measure the mean signal of all voids \citep[see e.g.][]{vielzeuf}. Knowing the angular size of voids, we re-scale the local mass map patches around the void centres. In such re-scaled units, we then extract convergence $\kappa$ patches five times the $R/R_{v}=1$ void radius, stack them to increase signal-to-noise, and measure radial profiles from the average $\kappa$ patch. Without a large set of simulations to estimate covariance of the void profile statistic, we estimate uncertainty using a void-by-void jackknife method \citep[see e.g.][]{Sanchez2017}. We then correct these re-sampling based uncertainties with reference to previous DES Y1 void analysis results that used more accurate Monte Carlo simulations \citep{vielzeuf}.

\begin{figure}
\begin{center}
\includegraphics[width=86mm
]{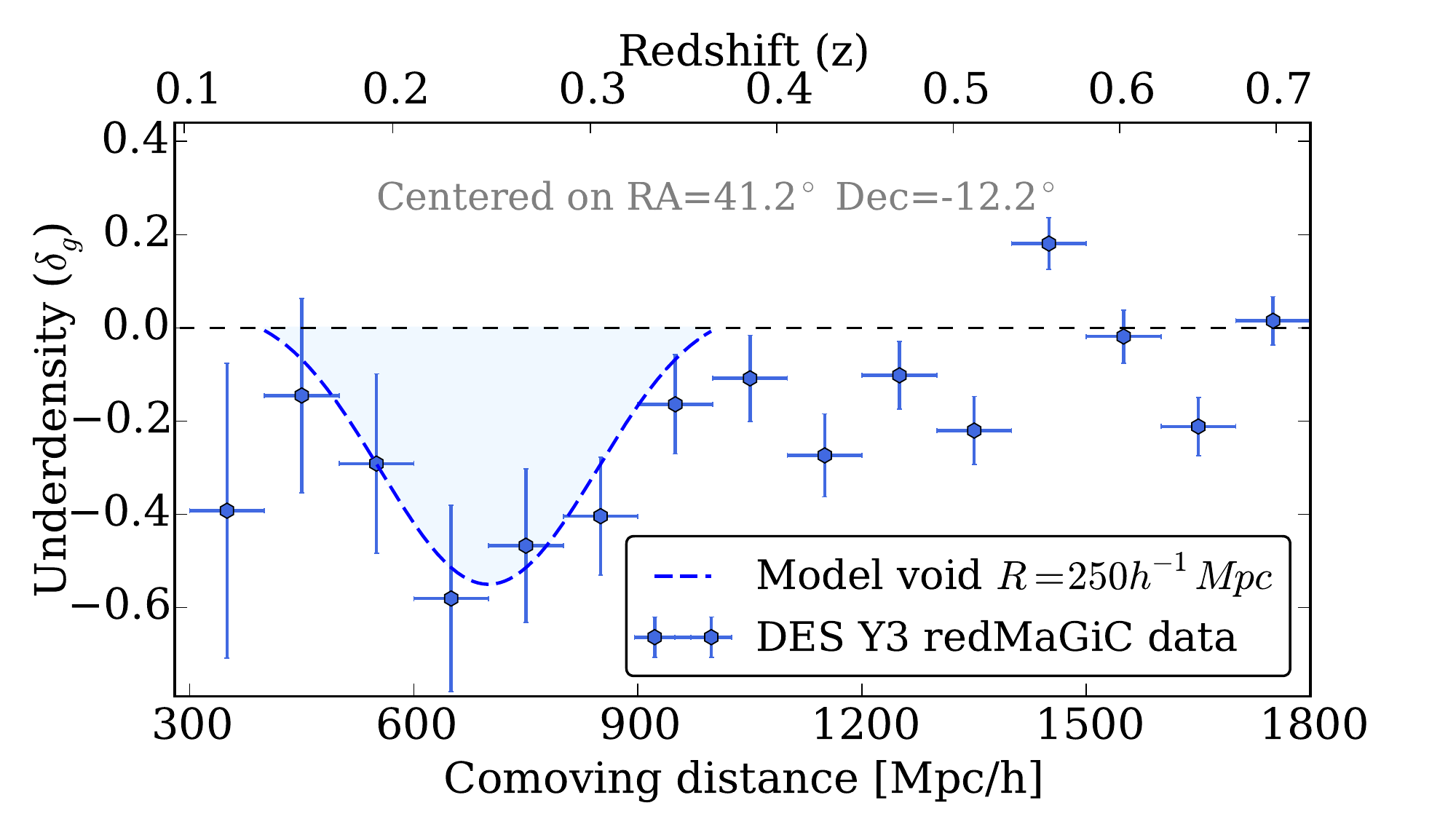}
\end{center}
\caption{The line-of-sight density of \redmagic{} galaxies aligned with a significant ``vole'' in the KS mass map. At low redshifts, we find evidence for an extended system of underdensities.   }
\label{fig:profile_LOS}
\end{figure}

Fig.~\ref{fig:void_profile} shows the measured profiles using the DES voids. As anticipated, we detect a negative convergence signal within the void radius ($R/R_{v}<1$) and a surrounding ring ($1<R/R_{v}<3$) of positive convergence signal (due to compensating mass around voids). We note that different mass map versions show consistent signals (within the quoted uncertainties). While these void lensing results remain open to much further quantitative work, there is certainly clear detection of correlations between underdensities of galaxies and matter; this will motivate further studies using DES Y3. We finally remark that the typical convergence signal associated with local underdensities can be affected by the void definition and selection. We explore alternative void samples extracted from DES Y3 data in Appendix~\ref{sec:AppVoids}.

\subsubsection{Line-of-sight underdensities}

Posing a slightly different question, we also examine the distribution of galaxies in a line-of-sight aligned with the most negative fluctuations in the DES Y3 mass maps. We call these \textit{voids in lensing maps} or \emph{voles} \citep[see e.g.][]{Davies2018}. We use a slightly modified version of the 2D void finder algorithm to identify them in the DES mass maps. We apply a Gaussian smoothing of 2~deg in order to intentionally select relatively deep and extended voles.

Following the previous DES Year 1 (Y1) analyses \citep{Chang2018}, the \redmagic{} galaxy position catalogue is projected into two-dimensional slices of 100~$h^{-1}~\textrm{Mpc}$ along the line-of-sight. This thickness corresponds to the approximate photo-$z$ errors of the \redmagic{} galaxies that allows the robust identification of voids \citep[see][for details]{Sanchez2017}. At redshifts $0.1<z<0.7$, galaxy density contrasts are measured in 15 tomographic slices aligned with voles. Galaxies are counted within an aperture of 2~deg of the void centre, which approximately corresponds to the full angular size of voles. The measured density contrasts at the different redshifts are  used to reconstruct the radial density profile aligned with the given vole. Fig.~\ref{fig:profile_LOS} shows the line-of-sight galaxy density aligned with a significant vole at $(\textrm{RA}, \textrm{Dec}) \approx (41.2^{\circ}, -12.2^{\circ})$ in the KS map.

We find an extended underdensity that is consistent with a super-void with radius $R_{\rm v}\approx250~h^{-1}~\mathrm{Mpc}$ (assuming simple Gaussian void profiles as in \citealt{Finelli2016}). This super-void, similar to the biggest underdensity found in the preceding DES Y1 analysis \citep{Chang2018}, will have smaller-scale substructures that are inaccessible using \redmagic{} photometric redshift data. Nevertheless, such a super-void is comparable to the largest known underdensities in the local Universe, and these objects are of great interest in cosmology \citep[see e.g.][]{Shimakawa2021}. Their integrated Sachs-Wolfe imprint has already been studied using DES Y3 data to probe dark energy \citep[for details see][]{Kovacs2019}.}

\section{Summary}
\label{sec:summary}

In this work we constructed weak lensing  convergence  maps  (`mass  maps')  from  the  DES Y3 data set using four reconstruction methods. The first method considered is the direct inversion of the shear field, also known as the Kaiser-Squires method, followed by a smoothing of small angular scales. The second method uses a prior on the B-modes of the map, imposing that the reconstructed convergence field must be purely an E-mode map (null B-mode prior); this method also includes smoothing at small scales. The third method, the Wiener filter, uses a Gaussian prior distribution for the underlying convergence field. Lastly, the \glimpse{} method implements a sparsity prior in wavelet (starlet) space, which can be interpreted as a physical model where the matter field is composed of a superposition of spherically symmetric haloes.

All methods are implemented on the sphere to accommodate the large sky  coverage  of  the  DES  Y3  footprint.  We  compared  the  different methods  using  simulations that are closely matched to the DES Y3 data. We quantified the performance of the methods at the map level using a number of different summary statistics: the Pearson coefficient with the `true' simulated convergence map, the root-mean-square error (RMSE) of the residual maps, the power spectra of the mass maps and residual maps, and the 1-point distribution function (PDF) of the mass maps.

The tests performed suggested that using our physically-motivated priors to recover the convergence field from a noisy realization of the shear field generally improves some aspects of the reconstruction. In particular, null B-mode, Wiener, and \glimpse{} delivered larger values of the Pearson coefficient and smaller values of the RMSE compared to the standard KS method, indicating that their use of physically-motivated informative priors significantly improve the accuracy of the reconstruction. We furthermore showed that a null B-mode prior mitigates the troublesome effects of masks and missing data. We also note how the choice of the prior can make the comparison of certain statistics with theoretical predictions non-trivial when taking the \textit{maximum a posteriori} result as a point estimate $\hat{\boldsymbol{\kappa}}$, rather than evaluating the full posterior distribution $p(\boldsymbol{\kappa}$). {Even if the effect of the prior cannot be easily modelled for a given theoretical summary statistic for cosmological inference, a forward modelling framework can be implemented that compares observed and simulated summary statistics.}

We have presented the official DES Y3 mass maps, obtained with the four different methods, and assessed their robustness against a number of systematic error maps representing catalogue properties and observing conditions. {  This recovered mass map, of which the dominant mass contribution is dark matter, covers the largest sky fraction of any galaxy weak lensing map of the late Universe.}

We emphasize that the choice of the particular mass map method depends on the goals and details of the science application. Science applications of these DES Y3 mass maps are expected in future work.

\section*{Acknowledgments}

Funding for the DES Projects has been provided by the U.S. Department of Energy, the U.S. National Science Foundation, the Ministry of Science and Education of Spain, 
the Science and Technology Facilities Council of the United Kingdom, the Higher Education Funding Council for England, the National Center for Supercomputing 
Applications at the University of Illinois at Urbana-Champaign, the Kavli Institute of Cosmological Physics at the University of Chicago, 
the Center for Cosmology and Astro-Particle Physics at the Ohio State University,
the Mitchell Institute for Fundamental Physics and Astronomy at Texas A\&M University, Financiadora de Estudos e Projetos, 
Funda{\c c}{\~a}o Carlos Chagas Filho de Amparo {\`a} Pesquisa do Estado do Rio de Janeiro, Conselho Nacional de Desenvolvimento Cient{\'i}fico e Tecnol{\'o}gico and 
the Minist{\'e}rio da Ci{\^e}ncia, Tecnologia e Inova{\c c}{\~a}o, the Deutsche Forschungsgemeinschaft and the Collaborating Institutions in the Dark Energy Survey. 

The Collaborating Institutions are Argonne National Laboratory, the University of California at Santa Cruz, the University of Cambridge, Centro de Investigaciones Energ{\'e}ticas, 
Medioambientales y Tecnol{\'o}gicas-Madrid, the University of Chicago, University College London, the DES-Brazil Consortium, the University of Edinburgh, 
the Eidgen{\"o}ssische Technische Hochschule (ETH) Z{\"u}rich, 
Fermi National Accelerator Laboratory, the University of Illinois at Urbana-Champaign, the Institut de Ci{\`e}ncies de l'Espai (IEEC/CSIC), 
the Institut de F{\'i}sica d'Altes Energies, Lawrence Berkeley National Laboratory, the Ludwig-Maximilians Universit{\"a}t M{\"u}nchen and the associated Excellence Cluster Universe, 
the University of Michigan, NFS's NOIRLab, the University of Nottingham, The Ohio State University, the University of Pennsylvania, the University of Portsmouth, 
SLAC National Accelerator Laboratory, Stanford University, the University of Sussex, Texas A\&M University, and the OzDES Membership Consortium.

Based in part on observations at Cerro Tololo Inter-American Observatory at NSF's NOIRLab (NOIRLab Prop. ID 2012B-0001; PI: J. Frieman), which is managed by the Association of Universities for Research in Astronomy (AURA) under a cooperative agreement with the National Science Foundation.

The DES data management system is supported by the National Science Foundation under Grant Numbers AST-1138766 and AST-1536171.
The DES participants from Spanish institutions are partially supported by MICINN under grants ESP2017-89838, PGC2018-094773, PGC2018-102021, SEV-2016-0588, SEV-2016-0597, and MDM-2015-0509, some of which include ERDF funds from the European Union. IFAE is partially funded by the CERCA program of the Generalitat de Catalunya.
Research leading to these results has received funding from the European Research
Council under the European Union's Seventh Framework Program (FP7/2007-2013) including ERC grant agreements 240672, 291329, and 306478.
We  acknowledge support from the Brazilian Instituto Nacional de Ci\^encia
e Tecnologia (INCT) do e-Universo (CNPq grant 465376/2014-2).

This manuscript has been authored by Fermi Research Alliance, LLC under Contract No. DE-AC02-07CH11359 with the U.S. Department of Energy, Office of Science, Office of High Energy Physics.

NJ has been supported by funding from l'Ecole Normale Sup\'erieure, Paris. OL and NJ acknowledge support from a European Research Council Advanced Grant TESTDE (FP7/291329) and STFC Consolidated Grants ST/M001334/1 and ST/R000476/1. AK has been supported by a Juan de la Cierva fellowship from MINECO with project number IJC2018-037730-I, and funding for this project was also  available in part through SEV-2015-0548 and AYA2017-89891-P. Cosmic voids computational work has been performed on the UK SCIAMA High Performance Computing cluster supported by the ICG, SEPNet and the University of Portsmouth.

\section*{Data availability}
The full metacalibration catalogue and mass maps will be made publicly available following publication, at \url{https://des.ncsa.illinois.edu}. 

\bibliographystyle{mnras}
\bibliography{bibliography,bibdesy3}
\appendix

\section{Author Affiliations}
\label{sec:affiliations}
{\small
$^{1}$ Laboratoire de Physique de l'Ecole Normale Sup\'erieure, ENS, Universit\'e PSL, CNRS, Sorbonne Universit\'e, Universit\'e de Paris, Paris, France\\
$^{2}$ Department of Physics \& Astronomy, University College London, Gower Street, London, WC1E 6BT, UK\\
$^{3}$ Institut de F\'{\i}sica d'Altes Energies (IFAE), The Barcelona Institute of Science and Technology, Campus UAB, 08193 Bellaterra (Barcelona) Spain\\
$^{4}$ Department of Physics and Astronomy, University of Pennsylvania, Philadelphia, PA 19104, USA\\
$^{5}$ Department of Astronomy and Astrophysics, University of Chicago, Chicago, IL 60637, USA\\
$^{6}$ Kavli Institute for Cosmological Physics, University of Chicago, Chicago, IL 60637, USA\\
$^{7}$ Instituto de Astrof\'{\i}sica de Canarias (IAC), Calle V\'{\i}a L\'{a}ctea, E-38200, La Laguna, Tenerife, Spain\\
$^{8}$ Departamento de Astrof\'{\i}sica, Universidad de La Laguna (ULL), E-38206, La Laguna, Tenerife, Spain\\
$^{9}$ Universit\"ats-Sternwarte, Fakult\"at f\"ur Physik, Ludwig-Maximilians Universit\"at M\"unchen, Scheinerstr. 1, 81679 M\"unchen, Germany\\
$^{10}$ Institute of Cosmology and Gravitation, University of Portsmouth, Portsmouth, PO1 3FX, UK\\
$^{11}$ Department of Physics, ETH Zurich, Wolfgang-Pauli-Strasse 16, CH-8093 Zurich, Switzerland\\
$^{12}$ AIM, CEA, CNRS, Universit\'e Paris-Saclay, Universit\'e de Paris, F-91191 Gif-sur-Yvette, France\\
$^{13}$ Argonne National Laboratory, 9700 South Cass Avenue, Lemont, IL 60439, USA\\
$^{14}$ Kavli Institute for Particle Astrophysics \& Cosmology, P. O. Box 2450, Stanford University, Stanford, CA 94305, USA\\
$^{15}$ Physics Department, 2320 Chamberlin Hall, University of Wisconsin-Madison, 1150 University Avenue Madison, WI  53706-1390\\
$^{16}$ Department of Physics, Carnegie Mellon University, Pittsburgh, Pennsylvania 15312, USA\\
$^{17}$ Instituto de Astrofisica de Canarias, E-38205 La Laguna, Tenerife, Spain\\
$^{18}$ Laborat\'orio Interinstitucional de e-Astronomia - LIneA, Rua Gal. Jos\'e Cristino 77, Rio de Janeiro, RJ - 20921-400, Brazil\\
$^{19}$ Universidad de La Laguna, Dpto. Astrofísica, E-38206 La Laguna, Tenerife, Spain\\
$^{20}$ Center for Astrophysical Surveys, National Center for Supercomputing Applications, 1205 West Clark St., Urbana, IL 61801, USA\\
$^{21}$ Department of Astronomy, University of Illinois at Urbana-Champaign, 1002 W. Green Street, Urbana, IL 61801, USA\\
$^{22}$ Department of Physics, Duke University Durham, NC 27708, USA\\
$^{23}$ Center for Cosmology and Astro-Particle Physics, The Ohio State University, Columbus, OH 43210, USA\\
$^{24}$ Jodrell Bank Center for Astrophysics, School of Physics and Astronomy, University of Manchester, Oxford Road, Manchester, M13 9PL, UK\\
$^{25}$ Department of Astronomy, University of California, Berkeley,  501 Campbell Hall, Berkeley, CA 94720, USA\\
$^{26}$ Santa Cruz Institute for Particle Physics, Santa Cruz, CA 95064, USA\\
$^{27}$ Fermi National Accelerator Laboratory, P. O. Box 500, Batavia, IL 60510, USA\\
$^{28}$ Department of Physics, The Ohio State University, Columbus, OH 43210, USA\\
$^{29}$ Jet Propulsion Laboratory, California Institute of Technology, 4800 Oak Grove Dr., Pasadena, CA 91109, USA\\
$^{30}$ Department of Physics, Stanford University, 382 Via Pueblo Mall, Stanford, CA 94305, USA\\
$^{31}$ SLAC National Accelerator Laboratory, Menlo Park, CA 94025, USA\\
$^{32}$ Department of Physics, University of Oxford, Denys Wilkinson Building, Keble Road, Oxford OX1 3RH, UK\\
$^{33}$ Department of Astronomy, University of Geneva, ch. d'\'Ecogia 16, CH-1290 Versoix, Switzerland\\
$^{34}$ Department of Physics, University of Michigan, Ann Arbor, MI 48109, USA\\
$^{35}$ Department of Applied Mathematics and Theoretical Physics, University of Cambridge, Cambridge CB3 0WA, UK\\
$^{36}$ Instituto de F\'isica Gleb Wataghin, Universidade Estadual de Campinas, 13083-859, Campinas, SP, Brazil\\
$^{37}$ Centro de Investigaciones Energ\'eticas, Medioambientales y Tecnol\'ogicas (CIEMAT), Madrid, Spain\\
$^{38}$ Brookhaven National Laboratory, Bldg 510, Upton, NY 11973, USA\\
$^{39}$ Institut d'Estudis Espacials de Catalunya (IEEC), 08034 Barcelona, Spain\\
$^{40}$ Institute of Space Sciences (ICE, CSIC),  Campus UAB, Carrer de Can Magrans, s/n,  08193 Barcelona, Spain\\
$^{41}$ Max Planck Institute for Extraterrestrial Physics, Giessenbachstrasse, 85748 Garching, Germany\\
$^{42}$ Institute for Astronomy, University of Edinburgh, Edinburgh EH9 3HJ, UK\\
$^{43}$ Cerro Tololo Inter-American Observatory, NSF's National Optical-Infrared Astronomy Research Laboratory, Casilla 603, La Serena, Chile\\
$^{44}$ Departamento de F\'isica Matem\'atica, Instituto de F\'isica, Universidade de S\~ao Paulo, CP 66318, S\~ao Paulo, SP, 05314-970, Brazil\\
$^{45}$ Instituto de F\'{i}sica Te\'orica, Universidade Estadual Paulista, S\~ao Paulo, Brazil\\
$^{46}$ CNRS, UMR 7095, Institut d'Astrophysique de Paris, F-75014, Paris, France\\
$^{47}$ Sorbonne Universit\'es, UPMC Univ Paris 06, UMR 7095, Institut d'Astrophysique de Paris, F-75014, Paris, France\\
$^{48}$ Department of Physics and Astronomy, Pevensey Building, University of Sussex, Brighton, BN1 9QH, UK\\
$^{49}$ University of Nottingham, School of Physics and Astronomy, Nottingham NG7 2RD, UK\\
$^{50}$ Astronomy Unit, Department of Physics, University of Trieste, via Tiepolo 11, I-34131 Trieste, Italy\\
$^{51}$ INAF-Osservatorio Astronomico di Trieste, via G. B. Tiepolo 11, I-34143 Trieste, Italy\\
$^{52}$ Institute for Fundamental Physics of the Universe, Via Beirut 2, 34014 Trieste, Italy\\
$^{53}$ Observat\'orio Nacional, Rua Gal. Jos\'e Cristino 77, Rio de Janeiro, RJ - 20921-400, Brazil\\
$^{54}$ Department of Physics, IIT Hyderabad, Kandi, Telangana 502285, India\\
$^{55}$ Faculty of Physics, Ludwig-Maximilians-Universit\"at, Scheinerstr. 1, 81679 Munich, Germany\\
$^{56}$ Institute of Theoretical Astrophysics, University of Oslo. P.O. Box 1029 Blindern, NO-0315 Oslo, Norway\\
$^{57}$ Instituto de Fisica Teorica UAM/CSIC, Universidad Autonoma de Madrid, 28049 Madrid, Spain\\
$^{58}$ Department of Astronomy, University of Michigan, Ann Arbor, MI 48109, USA\\
$^{59}$ Institute of Astronomy, University of Cambridge, Madingley Road, Cambridge CB3 0HA, UK\\
$^{60}$ Kavli Institute for Cosmology, University of Cambridge, Madingley Road, Cambridge CB3 0HA, UK\\
$^{61}$ School of Mathematics and Physics, University of Queensland,  Brisbane, QLD 4072, Australia\\
$^{62}$ Center for Astrophysics $\vert$ Harvard \& Smithsonian, 60 Garden Street, Cambridge, MA 02138, USA\\
$^{63}$ George P. and Cynthia Woods Mitchell Institute for Fundamental Physics and Astronomy, and Department of Physics and Astronomy, Texas A\&M University, College Station, TX 77843,  USA\\
$^{64}$ Department of Astrophysical Sciences, Princeton University, Peyton Hall, Princeton, NJ 08544, USA\\
$^{65}$ Instituci\'o Catalana de Recerca i Estudis Avan\c{c}ats, E-08010 Barcelona, Spain\\
$^{66}$ School of Physics and Astronomy, University of Southampton,  Southampton, SO17 1BJ, UK\\
$^{67}$ Computer Science and Mathematics Division, Oak Ridge National Laboratory, Oak Ridge, TN 37831\\
}

{ 
\section{Alternative samples of voids}
\label{sec:AppVoids}
We considered alternative catalogues of voids to test how the mass map imprints may depend on the void definition and selection. 

\texttt{VIDE}\footnote{https://bitbucket.org/cosmicvoids/} \citep{videpaper} is a watershed void finder based on \texttt{ZOBOV} \citep{neyrinck} that has been widely employed for various void studies \citep[see e.g.][and references therein]{Hamaus2020}. It has already been successfully used to study voids in the DES Y1 data \citep{Pollina2019, Fang2019}.

\texttt{VIDE}'s default centre is the volume-weighted barycentre, which does not generally coincide with the density minimum inside the void due to non-spherical void geometry. Instead, the barycentre preserves information about the void boundary. Therefore, a different kind of imprint signal is expected when correlated with convergence maps, with more pronounced positive rings rather than negative centres \citep[for a comprehensive study on the $\kappa$ signal associated with voids see][]{Cautun2016}. In the DES Y3 \redmagic{} data, \texttt{VIDE} detected $12,841$ voids. We then halved this catalogue using the \textit{compensation} of voids to further increase and isolate the expected signal from the boundary zone, expecting to see an enhanced positive convergence $\kappa$ imprint from these over-compensated voids.

\begin{figure}
\begin{center}
\includegraphics[width=86mm]{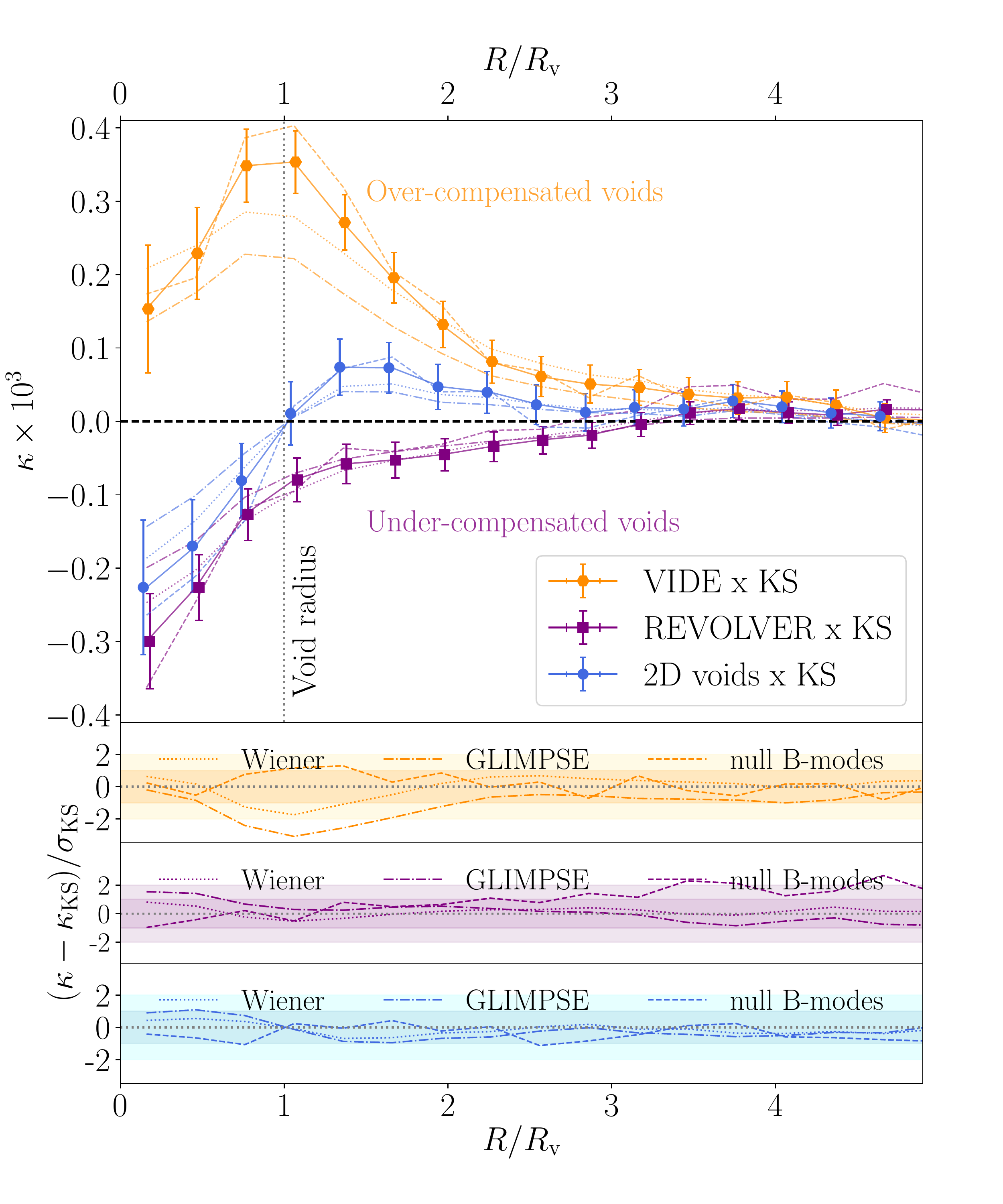}
\end{center}
\caption{\textit{Top panel:} Different mass map imprints of different types of voids. \textit{Bottom panel:} Differences in signals measured from different mass map reconstructions, relative to the KS results and errors (shaded ranges are $1\sigma$ and $2\sigma$ around the KS results).  }
\label{fig:void_profile_more}
\end{figure}

We are also interested in detecting the most pronounced negative $\kappa$ signals associated with a specific subclass of large and deep voids that are \emph{under}-compensated. As a third option, we thus used the public\footnote{https://github.com/seshnadathur/REVOLVER/} void finder algorithm \texttt{REVOLVER} \citep{Nadat2018,beyondbao}, also based on the \texttt{ZOBOV} algorithm.

A proxy for the gravitational potential (and thus for the convergence field) at the positions of voids can be defined as
\begin{equation}
\label{eq:lambda_v}
\lambda_v\equiv\overline\delta_g\left(\frac{R_{\rmn{eff}}}{1\;h^{-1}\rmn{Mpc}}\right)^{1.2} \ \ ,
\end{equation}
using the average galaxy density contrast $\overline\delta_g = \frac{1}{V}\int_{V}\delta_g\,\rmn{d}^3\mathbf{x}$ and the effective spherical radius, $R_\rmn{eff}= \left(\frac{3}{4\pi}V\right)^{1/3}$, where the volume $V$ is the total volume of the void \citep[for further details see][]{NadathurCrittenden2016,Nadathur2017}. 
\cite{cmbboss} showed that different values of the $\lambda_{v}$ parameter indicate different (CMB) lensing imprints, including signals with either positive or negative sign, aligned with the void centre\footnote{\texttt{REVOLVER} voids may also be defined using barycentres.}. Following this, we keep only $7,782$ of  the most under-compensated voids defined by the lowest $\lambda_{v}$ values. Leaving more detailed analyses for future work, we note that a subclass of voids with high $\lambda_{v}$ values would also correspond to over-compensated voids such as our \texttt{VIDE} sample.

Fig.~\ref{fig:void_profile_more} shows the measured profiles of our \texttt{REVOLVER}, \texttt{VIDE}, and 2D void analyses given the uncertainties. As anticipated based on the differences in the nature of the voids we selected, we detected qualitatively different signals in each case:
\begin{itemize}
\item the \texttt{VIDE} voids show a relative depression in convergence at the void centre compared to the pronounced peak at the void boundary, matching our expectations.
\item the \texttt{REVOLVER} voids we selected are associated with strong negative $\kappa$ imprints that in fact extend far beyond the void radius, indicating surrounding voids on average.
\item 2D voids combine the advantages of the other finders. They excel in marking the actual radius of voids in the mass map profiles, with reduced central and wall amplitudes.
\end{itemize}

We thus report that all three void types we consider show consistent signals when mass maps are varied for a given void sample. We leave more detailed analysis for future work.}

\end{document}